\documentclass[pra,twocolumn,amsmath,amssymb,groupedaddress,superscriptaddress]{revtex4}
\usepackage{graphics}
\usepackage{amsmath}
\usepackage[dvips]{graphicx}
\usepackage{float,graphicx}

\def\>{\rangle}
\def\<{\langle}

\def\be{\begin{eqnarray}}
\def\beqa{\begin{eqnarray}}

\def\ee{\end{eqnarray}}
\def\eeqa{\end{eqnarray}}

\def\aat{\textmd{at}}

\def\eeff{\textmd{eff}}
\def\iint{\textmd{int}}

\def\mmax{\textmd{m}}

\def\ddiss{\textmd{diss}}
\def\iint{\textmd{int}}

\def\DDOS{\textmd{DOS}}
\def\ttot{\textmd{tot}}
\def\ttd{\textmd{3D}}
\def\ddiss{\textmd{diss}}

\def\Ttr{\textmd{Tr}}
\def\aat{\textmd{at}}

\def\ppol{\textmd{pol}}
\def\Rre{\textmd{Re}}

\def\eenv{\textmd{env}}
\def\llamb{\textmd{LS}}
\def\be{\begin{equation}}
\def\ee{\end{equation}}
\def\bea{\begin{eqnarray}}
\def\eea{\end{eqnarray}}
\usepackage{bm}
\usepackage{graphicx}
\begin{document}
\title[Lattice mapping for many-body open quantum systems]{Lattice mapping for many-body open quantum systems and its application to atoms in photonic cystals}
\author{In{\'e}s de Vega}
\affiliation{Department of Physics and Arnold Sommerfeld Center for Theoretical Physics, Ludwig-Maximilians-Universit{\"a}t M{\"u}nchen, Theresienstr. 37, 80333 Munich, Germany}
%S.F. Huelga$^{\dag}$ and $^{\dag \dag}$}
%\affiliation{$^{\gamma}$Departamento de F\'{\i}sica Fundamental II. Universidad de La Laguna,
%La Laguna 38203, Tenerife, Spain}
%\affiliation{$^{\ddag}$Departamento de F\'{\i}sica Fundamental
%y Experimental, Electr\'onica y Sistemas. Universidad de La Laguna,
%La Laguna 38203, Tenerife, Spain}
%\affiliation{$^{\dag}$Center for nonlinear phenomena and complex systems.
%Universit\'e Libre de Bruxelles. Bvd du Triomphe 1050-Bruxelles}
%\affiliation{$^{\dag\dag}$Physikalisches Institut,
%Universit\"at Freiburg, Hermann-Herder-Str. 3, 79104-Freiburg, Germany}
%%%%%%%%%%%%%%%%%%%%%%%%%%%%%%%%%%%%%%%%%%%%%%%%%%%%%%%%%%%%%%%%%%%%%%%%%
\begin{abstract}
We present a derivation that maps the original problem of a many body open quantum system (OQS) coupled to a harmonic oscillator reservoir into that of a many body OQS coupled to a lattice of harmonic oscillators. The present method is particularly suitable to analyze the dynamics of atoms arranged in a periodic structure and coupled the EM field within a photonic crystal. It allows to solve the dynamics of a many body OQS with methods alternative to the commonly used master, stochastic Schr\"o{}dinger and Heisenberg equations, and thus to reach regimes well beyond the weak coupling and Born-Markov approximations. 
\end{abstract}
%\tableofcontents
\maketitle
% Motivation
The analysis of matter at nano-scales is not only of fundamental interest, but it is also of primary importance to develop future technologies, and to create new materials and devices with a vast range of applications. Indeed, nanostructures may display intriguing quantum phenomena, 
%nowadays measured sophisticated experiments that enables to solve the system dynamics at fembto second scales, like for instance two photon spectroscopy used to analyzed quantum beats in the energy transport of photosynthetic complexes. These structures 
that nowadays can be accurately controlled in experimental setups.
% such as quantum cavities, atomic lattices, atomic ensembles or trapped ions. In some of these controlled systems, the quantum phenomena of interest occur within time scales where the effects of the environment can be neglected. 
However, in most cases, quantum systems shall be considered as {\bf open} (OQS)\cite{breuerbook,rivas2011a}, i.e. interacting with environments such as the radiation field or phonons within a lattice. In addition, in order to describe such interaction, it is often not accurate to assume a large separation of scales between the quantum system and its environment, like in the weak coupling or Markov approximations, or to assume that the relevant part of the system can be described in a reduced subspace of the total Hilbert space, like in the so-called projection operator techniques %and hence a description of the system beyond the weak coupling and the so-called Markov approximation 
\cite{nakajima1958,zwanzig1960,breuer2004,breuer2007,breuerbook,esposito2003,budini2005}. Hence, an alternative approach that allows to go beyond these assumptions is highly desirable.  

%Our increasing ability to fabricate new materials, and to observe and control quantum systems at different times, length scales and energy ranges is constantly revealing new scenarios where dissipation and decoherence play a fundamental role. In many of these scenarios a large separation between system and environment time scales can no longer be assumed, what leads to a non-Markovian behavior and a eventual back-flow of information from the environment into the system. It is therefore crucial to develop an accurate but efficient description of the system-environment interaction that goes well beyond the Born-Markov approximation. 

Of particular interest are {\bf many body} OQSs, such as impurities in a solid or molecules in a photosynthetic complex. In some cases, such particles are strongly correlated, leading to a rich behaviour.  To analyse such strongly correlated systems,   %of their intrinsic interest, as well as their many applications, 
powerful numerical tools, like for instance density matrix re-normalization methods (DMRG) \cite{white1992,white1998,ostlund1995}, as well as advanced experimental techniques, like atomic lattices \cite{bloch2005} and trapped ions \cite{leibfried2003,britton2012}, have been developed. Nevertheless, the study of strongly correlated systems coupled to an environment is still in its initial stages. The interplay between the many-particle correlations and dissipation has begun to be analysed recently \cite{barreiro2010,schindler2012,calzadilla2006,prosen2008,daley2009,schwanger2013,cai2013}, but always within the Markov approximation.
Hence, extending such analysis to non-Markovian and strong coupling interactions is still an open problem that may give rise to a wealth of novel phenomena and applications.

% Comparison previous approaches
To describe an OQS beyond the weak coupling regime, 
%one possibility is to integrate the dynamics of the total system, diagonalizing the full Hamiltonian (add references). An alternative is to re-express the whole system in such a way that it is exactly tractable with state of the art techniques. Based on the last idea, 
one possibility is to re-express the whole system in such a way that it can be exactly tractable with state of the art techniques. Based on this idea is the chain representation proposed in \cite{bulla2008,hughes2009,prior2010,chin2010}, which consists on performing a unitary transformation on the environment to re-express the full system as the OQS coupled to a linear chain of modified harmonic oscillators. After this mapping, it becomes possible to simulate the unitary evolution of the total system wave function $|\Psi (t)\rangle$ \cite{prior2010}.
%Chain representations have been known for quite some time in both classical and quantum contexts \cite{bulla2008,hughes2009}, but the approach \cite{prior2010,chin2010} use the theory of orthogonal polynomials to perform an exact and analytic mapping. 
This mapping is particularly useful for a single or few particles $N$ forming the OQS \cite{chin2013}, either connected each of them to independent environments, or connected to the same environment uninformly, i.e. with equal weight (see Fig.1(a)). 
%scale with the number of particles extensions for a higher number require interactions within the chain that may go far beyond next-neighbors, what complicates the numerical resolution of the problem. 

\begin{figure}[ht]
%\centerline{\includegraphics[width=0.5\textwidth]{chain2.pdf}}
%\centerline{\includegraphics[width=0.5\textwidth]{tresH.pdf}}
%\centerline{\includegraphics[width=0.5\textwidth]{ChainOnly.pdf}}
\centerline{\includegraphics[width=0.5\textwidth]{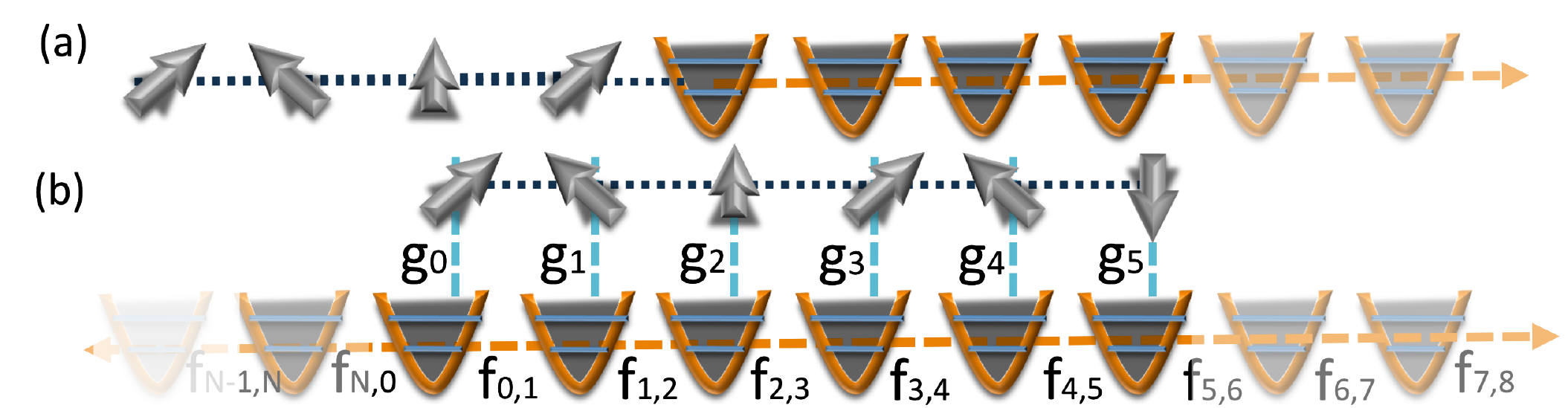}}
\caption{(Color online) (a) Several atoms uniformly coupled to the same environment, according to the proposal by \cite{chin2013}. (b) Ladder structure here proposed, corresponding to the mapped Hamiltonian (\ref{fintrans}). We have denoted $g_n=gu_n$. \label{statNM}}
\end{figure}

This paper proposes an alternative unitary transformation that maps the problem of a many body OQS coupled non-uniformly to a harmonic oscillator environment, into that of a many body OQS coupled \textit{in a ladder structure} to a $1D$ chain of transformed harmonic oscillators (See Fig.\ref{statNM}(b)). The present method is particularly suitable to analyse the dynamics of atoms arranged in a periodic structure and coupled the EM field within a photonic crystal structure having either one, two or three dimensional photonic band gaps \cite{yablonovitch1987,john1987}. In this regard, we show the performance of the mapping by analyzing the dynamics of atoms within a one-dimensional photonic crystals, and comparing this analysis with the results obtained by using a master equation. It is also shown how an exact diagonalization of the mapped system unveils the presence of polaritons, which are formed when the atomic frequencies are placed within the photonic crystal gap. These polaritons are highly correlated atom-photon states, and can be related to the existence of an incomplete atomic relaxation. 

Indeed, in a similar way to previous proposals, the mapped structure may facilitate solving the system dynamics and equilibrium properties by performing a systematic truncation of the environment and then evolving with a Schr\"o{}dinger equation, or by using traditional many body methods, like exact diagonalization, Montecarlo \cite{pollet2012}, DMRG or time-adaptive-DMRG (t-DMRG) \cite{vidal2003,scholl2011}, and the mean field approximation in the case of higher dimensional structures. This allows to reach regimes that go far beyond those that are well described with a master equation or a stochastic Schr\"o{}dinger equation \cite{breuerbook,carmichael1993}. In addition, the mapped system may present a similar structure as well known models, like the extended Jaynes-Cummings model appearing arrays of coupled cavities \cite{greentree2006,hartmann2006,angelakis2007}.
% that have been widely analysed with mean field theories). Also, the transformed Hamiltonian may be subject to further unitary transformations like the displaced-operator transformation, allowing to gain more insight into the problem. 
%It also allows to compare the results to those of master equations obtained with different approximations and assumptions, thus giving rise to an analysis of the accuracy of the last.
Finally, the mapped system can be implemented in atomic lattices, so that the interaction of an OQS with different environments may be experimentally engineered, analysed and measured.   
%{\bf One excitation sector--}

\section{The model}
Let us consider the following Hamiltonian,
 \beqa
H&=&H_S+\sum_k \omega(k) a_k^\dagger a_k+\sum^N_{n=1}g_n\int_{-k_s}^{k_s}dk u_n (a_k^\dagger \phi_n(k)\cr
&+& a_k \phi^*_n(k)) C_n.
\label{h1EM}
\eeqa
with $k_s$ the maximum wave vector $k$ of the environment, $N$ the number of particles in the quantum system, $a^\dagger_k (a_k)$ the creation (annihilation) operators corresponding to the environment mode $k$ with frequency $\omega(k)$, $u_n$ the coupling strength of the system particle $n$ with such field, and $C_n$ a system coupling operator corresponding to the particle $n$. Also, $H_S$ is the system free Hamiltonian, and the functions $\phi_n(k)$ are here assumed to form an orthonormal set with the property (i) $\int_{-k_s}^{k_s}dk\phi^*_n(k)\phi_m(k)=\rho_0\delta_{nm}$, with $\rho_0$ a normalization factor.
%, within a given interval $[-k_s,k_s]$.
%The former Hamiltonian corresponds to a many body open quantum system coupled to a vector field, such as the electromagnetic field. However, when the OQS couples to a scalar field, 
% Note that in some cases, the orthonormal functions need not to be complex. 
%\beqa
%H=H_S+\sum_k \omega(k) a_k^\dagger a_k+\sum_n\sum_k u_n \phi_n(k) a_{xk} C_n,
%\label{h0}
%\eeqa
%where $a_{xk}=a_k^\dagger+ a_k$, 
Let us now consider the transformation $b_n=\int_{-k_s}^{k_s}dk \phi^*_n(k)a_k$, with the inverse
$a_k=\sum_m \phi_m(k)b_m$. To ensure that the conmutation (or anticommutation) properties of the original operators are conserved in the transformed ones, $b_n$, %, such that $[b_n,b_m^\dagger]=\delta_{nm}$ for bosons, and this is achieved with the property and 
we need an additional property (ii) $\sum_n \phi^*_n(k)\phi_n(k')=\rho_0\delta(k-k')$. Properties (i)-(ii) guarantee that the transformation is unitary.
Applying this transformation to (\ref{h1EM}) we obtain
\beqa
H=H_S+\sum^M_{n,m=1}f_{nm} b_n^\dagger b_m+g\sum^N_{n=1} u_n (b_n^\dagger+b_n)C_n,
\label{fintrans}
\eeqa
where we have assumed that $g_n=gu_n$, and $f_{nm}=\int_{-k_s}^{k_s}dk \omega_k\phi^*_n(k)\phi_m(k)$ is a function that decays with $|n-m|$ depending on the particular dispersion relation, and the orthonormal basis choosen. The original Hamiltonian has been mapped to that of a chain of transformed modes $b_n$ coupled one by one to each particle $n$. Note that although the proposed transformation %(\ref{transfo}) r
may require a large number $M$ of modes $b_n$ to correctly map the environment, the number of particles $N$ in the OQS can in principle be arbitrary. %Indeed, $H_S$ may only be restricted to a few number of particles $N$.

The mapping is valid for both bosonic and fermionic reservoirs, and constitutes an interesting playground to analyze the dynamics of many body open quantum systems. 
%The mapped problem described with (\ref{fintrans}) can in principle be analyzed exactly with powerful numerical tools such as t-DMRG, and provides an extension for the many body case of the chain mapping methods developed in \cite{prior2010,chin2010,chinbook2011}. 
% The resulting system has a ladder structure which has been already analyzed in the context of finite temperature simulations with $DMRG$ ({\bf check more}) \cite{scholl2010,karrasch2013}.
Furthermore, it can be used to describe the dynamics of many physically realistic problems. To illustrate this, in the following we analyze how light-matter interaction Hamiltonians can be tailored to reach the desired form (\ref{h1EM}), from which the proposed mapping can be performed.
 
%{\bf Inhomogeneous couplings--}
%Let us consider a Hamiltonian of the form 
%\beqa
%H&=&H_S+\sum_k \omega(k) a_k^\dagger a_k+\sum^N_{n=1}\sum_k h_n(k)(a_k^\dagger \phi_n(k)\cr
%&+& a_k \phi^*_n(k)) C_n,
%\label{h1}
%\eeqa
%which differs from (\ref{h1EM}) because of the coupling term $h_n(k)=u_n g(k)$.

%\beqa
%H=H_S+\sum_k \hat{\omega}(k) a_k^\dagger a_k+\sum_n\sum_k u_n \phi_n(k) a_{xk} C_n,
%\label{equiv}
%\eeqa
%which clearly has the form (\ref{h1EM}).
\section{One dimensional electromagnetic fields}
Let us consider a Hamiltonian of the form 
\beqa
H&=&H_S+\sum_k \omega(k)a_k^\dagger a_k+\sum^N_{n=1}\sum_k g_n(k)(a_k^\dagger u_k(r_n)\cr
&+& a_k u^*_k(r_n)) C_n,
\label{h1}
\eeqa
with 
\begin{eqnarray}
g_{n}(k)=-i \sqrt{\frac{1}{2\hbar \omega(k) \epsilon_0\nu}}\omega_{n}\sum_\sigma\hat{e}_{{\bf k},\sigma}\cdot{\bf d}^n_{12},
\label{chapuno23}
\eea
where $\nu$ is the quantization volume, $\epsilon_0$ is the free space permitivity, and ${\bf d}^n_{12}=d^n_{12}{\bf u}_d$ and $\omega_n$ are respectively the dipole moment and resonant frequency of the $n$-th atom. In the following, we assume for simplicity that all atoms have equal resonant energy $\omega_0$ and dipolar moment $d_{12}$. The quantity $\hat{e}_{{\bf k},\sigma}$, is the polarization vector corresponding to the wave vector ${\bf k}$ with polarization $\sigma$. Here we have assumed that the energy absorption and emission process is independent on the polarization of the photons, i.e. $a_{k,\sigma}\equiv a_k$. 
%Assuming that the wave vector propagates along a single direction, $\hat{e}_{{\bf k}}$ propagates perpendicularly to such direction.
\begin{figure}[ht]
%\centerline{\includegraphics[width=0.5\textwidth]{chain2.pdf}}
%\centerline{\includegraphics[width=0.5\textwidth]{tresH.pdf}}
%\centerline{\includegraphics[width=0.5\textwidth]{ChainOnly.pdf}}
\centerline{\includegraphics[width=0.5\textwidth]{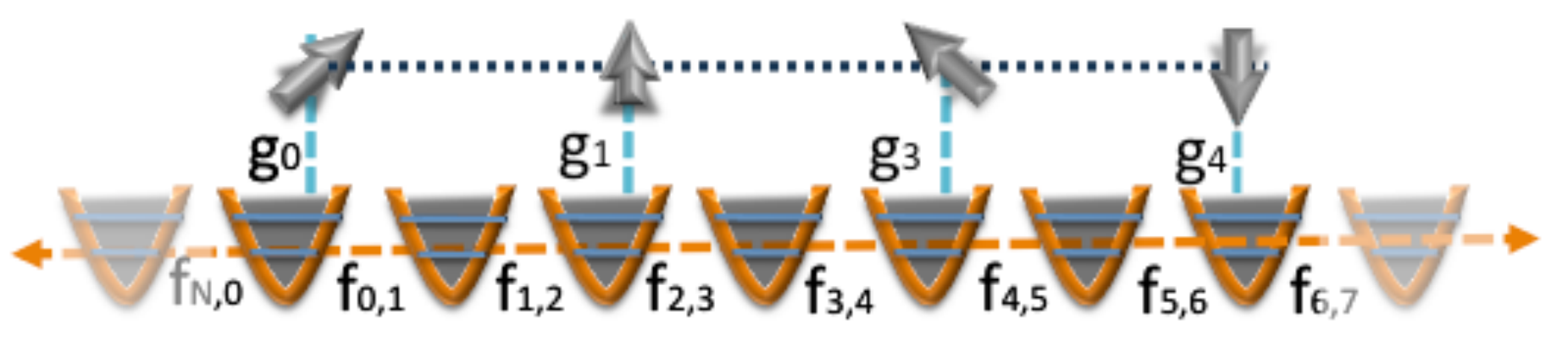}}
\caption{(Color online) Ladder structure for $P=2$. The fact that the atoms are separated by a distance $d_0=2h_0$ gives rise to a situation where only one every two sites in the harmonic oscillator lattice is connected with an atom. \label{statNM2}}
\end{figure}
The functions $u_k(r_n)$ are the so-called mode functions, that in the continuum form an orthonormal set within the physical region under consideration, i.e. $\int dr u^*_k(r)u_k(r)=\delta_{kk'}$. These functions are solutions of the wave equation, and depend on the boundary conditions considered: while periodic boundary conditions correspond to travelling-wave modes, reflecting walls lead to standing waves \cite{wallsbook}. For the earlier case, $u_k(r_n)=(1/L)e^{ikr_n}$, and the wave vector $k$ takes the values $k=2\pi q/L$, with $L=h_0M$ defining the physical quantization region, and $q=1,\cdots,M$. 
%\beqa
%H&=&H_S+\int^{k_\mmax}_{-k_\mmax} dk\omega(k) a_k^\dagger a_k+\sum^N_{n=1}\int^{k_\mmax}_{-k_\mmax} dk h_n(k)(a_k^\dagger \phi_n(k)\cr
%&+& a_k \phi^*_n(k)) C_n,
%\label{h1}
%\eeqa
%which differs from (\ref{h1EM}) because of the coupling term $h_n(k)=u_n g(k)$.
Naturally, to obtain this Hamiltonian from (\ref{h1EM}), we may just consider $\phi_n=h_0u_k(r_n)$, and assume that $g(k)\approx g$. 
%We shall consider the momentum discretized as $k_q=2\pi q/(Mh_0)$, for $q=1,\cdots,M$, and $Mh_0$ the physical length under consideration. 
In order to fulfill the relation (ii) $(1/M)\sum_n\phi^*_n(k)\phi_m(k')=\delta(k-k')$, we shall chose atoms with positions $r_n=nd_0$, i.e. spaced by a constant distance given by $d_0=Ph_0$, where $P=1,2,\cdots$. In this situation, the proposed transformation becomes simply a discrete Fourier transform, and the mapped Hamiltonian has the form 
\beqa
H=H_S+\sum^M_{n,m=1}f_{nm} b_n^\dagger b_m+\sum^N_{n=1} g_n (b_{nP}^\dagger+b_{nP})C_n,
\label{fintrans2}
\eeqa 
Fig. (\ref{statNM2}) represents this structure for $P=2$. The larger the separation between atoms, $P$, the more spacing between the harmonic oscillators that are coupled to atoms within the chain. 

%The above condition, also imply that $e^{ikr_n}=e^{ikr_{n+M}}$, but this 
%
%$\alpha_{nm}(t)=u_nu_m\int^{k_s}_{-k_s} dk \phi^*_n(k)\phi_m(k)e^{-i\hat{\omega}(k)t}$,
%is equal to $\alpha^{1D}_{nm}(t)$. Here $[-k_s,k_s]$ is the interval in which the functions $\phi_n(k)$ are orthogonal.
%%The integral in (\ref{h1corr}) ranges within an interval $[-k_\mmax,k_\mmax]$. 
%Assuming that $k_\mmax=\pi/d_0$, a wave vector large enough as to account for the most important part of the field spectrum. In that case, we can make in (\ref{h1corr}) a change of variable such that $k\rightarrow kd_0$, so that in the continuum limit we have 
%$\alpha^{1D}_{nm}(t)=\int^\pi_{-\pi} g^2(k) e^{ik(n-m)-i\omega(k)t}$. A natural choice is $\phi_n=(1/\sqrt{M})e^{i k n}$, so that $k_s=\pi$. To make sure that (ii) $(1/M)\sum_n\phi^*_n(k)\phi_m(k')=\delta(k-k')$ is fulfilled, we also need periodic boundary conditions for the OQS lattice i.e. $e^{ikr_n}=e^{ikr_{n+M}}$. In this case, considering $r_n=nd_0$, we find that $k=2\pi l/(Md_0)$. Notice that $Md_0$ is the size of the total system (i.e. the space where the EM field is defined), so that $\sum_{l=-M/2}^{M/2} e^{i 2\pi l |n-m|/M}=M\delta_{nm}$. 
Periodicity in the boundary conditions is not as restrictive as it seems: in general $N\ll M$, and atoms from $n=N+1$ to $n=M$ are just virtual particles introduced for the transformation. Hence, boundary effects do not affect drastically the systems dynamics, as long as the system is sufficiently small as compared to the environment.  

Interesting at this point is to realize that when the initial state of the environment is thermal, the so-called correlation function is the only quantity necessary to describe the dynamics of an OQS coupled to it \cite{chin2011}. An exact calculation of the Heisenberg evolution equations of the system operators \cite{alonso2007}, leads to the conclusion that indeed the correlation function is the only quantity needed to describe the effects of the environment in the system dynamics (see Appendix A for details). Also, the correlation function appears in master equations derived with different methods \cite{breuerbook}. Hence, replacing $g_n(k)\approx g_n$ (with $g=g(k_\eeff)$, and $k_\eeff$ is the resonant wave-vector, that can be defined as $\omega(k_\eeff)=\omega_0$) in (\ref{h1}) is a reasonable approximation when in the calculus of the correlation function
\bea
\alpha^{1D}_{nm}(t)=\frac{h_0}{2\pi}\int^{k_\mmax}_{-k_\mmax} dk g_n(k)g_m(k) e^{ikr_{nm}-i\omega(k)t},
\label{correl1D}
\eea
the term $\exp(i\omega(k)t)$ varies much faster in $k$ than $g(k)$, which can then be considered as a constant in comparison \cite{devega2005}. In other words, in many problems the structure of the environment is captured mainly by the dispersion relation, which in turn fully determines the density of states.
% (see for instance this approximation in Fig. (\ref{approx1}a-b) for an Ohmic reservoir, where $g^2(k)=k$). 
%\begin{figure}[ht]
%%\centerline{\includegraphics[width=0.45\textwidth]{ohmic2.pdf}}
%\centerline{\includegraphics[width=0.5\textwidth]{ohmictot.pdf}}
%\caption{Plots (a)-(b): Correlation function (\ref{h1corr2}) of an Ohmic reservoir, when considering $q^2(k)=k$ within the integral (solid curves), and outside the integral (dashed curves). Blue and orange colors represents respectively real and imaginary parts of the correlation function, and we have choosen $n=0$, so that Plot (a) corresponds to $m=0$ ((a)) and (b) to $m=1$; Plots (c)-(d): Decaying of coefficient with $|n|$ ($m=0$) in logarithmic scale, for the Ohmic (transparent circles), sub-Ohmic (dark circles) and super-Ohmic (squares) reservoirs of Example 2. \label{approx1}}
%\end{figure}
%When the environment is initially in a Gaussian state, it is possible to integrate out the environment degrees of freedom and derive a formally-exact expression for the reduced state dynamics of the atomic system using path integral methods \cite{leggett1987}. Also, it is possible to derive the set of Heisenberg equations that obey the system operators \cite{devega2006,alonso2007}. 
In the mapped Hamiltonian (\ref{fintrans}), in general $f_{nm}$ decays exponentially for low-order polynomial $\omega_k$, so provided that $\omega(k)$ is sufficiently smooth in $k$, $f_{nm}$ can be truncated at near or next-near neighbors. % (see Fig. (\ref{approx1}))
This truncation is exact for periodic $\omega_k$ as we will see in Section V. 

\section{Three dimensional EM fields}

When the many body open quantum system is connected to a vector field within a three dimensional space, the initial Hamiltonian may have a different form from (\ref{h1EM}). This is the case of atoms coupled to the electromagnetic field, where $H=H_S+H_B+H_\iint$, where $H_B=\sum_{\bf k}\omega({\bf k})a^\dagger_{\bf k}a_{\bf k}$. Considering the case of travelling waves, such that the magnitude of the mode functions is $u_{\bf k}({\bf r})=L^{-3/2}e^{i{\bf k}\cdot{\bf r}}$, the interaction Hamiltonian can be written as 
\beqa
H^\ttd_\iint=\sum_{\bf k}\sum_{\bf n}g({\bf k})\bigg(a_{\bf k} e^{i{\bf k}\cdot{\bf r_n}}+a^\dagger_{\bf k} e^{-i{\bf k}\cdot{\bf r_n}}\bigg)C_{\bf n},
\label{hEM1}
\eeqa
Here we have defined $g({\bf k})=-i \sqrt{\frac{1}{2\hbar \omega_{\bf k} \epsilon_0\nu}}\omega_{0}\sum_\sigma {\bf e}_{{\bf k},\sigma}\cdot{\bf d}_{12}$, considering for simplicity that every atom has the same frequency $\omega_0$. We now assume also that the atoms are placed in a cubic lattice separated by a distance $d_0$ in each direction, so that ${\bf r}_n=d_0 {\bf n}=d_0(n_x,n_y,n_z)$, and the sums in ${\bf n}$ go from $n_\beta=1,\cdots, N_\beta$, with $N_\beta$ the number of atoms considered in the direction $\beta=x,y,z$. The emission of atoms in such a regular crystal-like structure have also been analyzed in \cite{porras2008}. The wave vector ${\bf k}$ takes the values $k_\beta=2\pi q_\beta/L_\beta$, with $L_\beta=h_0M_\beta$ defining the physical quantization region in the direction $\beta$, and $q_\beta=1,\cdots,M_\beta$. For simplicity, we shall define $M_\beta=M$, and $N_\beta=N$, and $d_0=h_0$.
Let us consider in addition that $g({\bf k})\approx g(k)\approx g(k_\eeff)=g$, where $k_\eeff$ is the resonant wave-vector. As in the one dimensional case, the former is a good approximation when the phase $\exp(i\omega(k)\tau)$ varies much faster in ${\bf k}$ (both in module and angular dependency) than the coupling coefficient, so that
\bea
\alpha_{nm}(\tau)&=&\gamma(\frac{h_0}{2\pi})^3\sum_{\sigma}\int d{\bf k}\frac{ |\hat{e}_{{\bf k},\sigma}\cdot \hat{u}_d|^{2}}{\omega(k)}e^{i{\bf k}\cdot{\bf r}_{nm}-i\omega(k)\tau}\cr
&\approx&\gamma\int d{\bf k} e^{i{\bf k}\cdot{\bf r}_{nm}-i\omega(k)\tau}
\label{corr3D}
\eea
where $\gamma=\hat{\gamma}(\frac{h_0}{2 \pi})^3 \sum_{\sigma}\frac{ |\hat{e}_{k_\eeff,\sigma}\cdot \hat{u}_d|^{2}}{\omega(k_\eeff)}$, with $\hat{\gamma}=\omega_{0}^2 d^2_{21}/(2\hbar\epsilon_0\nu)$, is a good approximation to the correlation function. Note that the quantization volume $\nu=h_0^3$.

In the following, we shall consider two different cases: an anisotropic dispersion relation, and an isotropic one. 

\subsection{Anisotropic dispersion relations}
For an anisotropic dispersion relation, it is consider to assume a three dimensional Foruier transform $a_{\bf k}=\sum_{\bf m}e^{i{\bf k}\cdot {\bf r}_m}b_{\bf m}$, which leads to the simple form 
\bea
H=H_S+\sum^M_{\bf n,m=1}f_{\bf nm} b_{\bf n}^\dagger b_{\bf m}+g\sum_{\bf n} (b_{\bf n}^\dagger+b_{\bf n})C_{\bf n}.
\label{fintrans3D}
\eea
Here, the quantity
\bea
f_{\bf nm}=\sum_{\bf k}\omega({\bf k})e^{-i{\bf k}\cdot{\bf r}_{nm}}
\eea
with ${\bf r}_{nm}={\bf r}_n-{\bf r}_m$. Notice that for anisotropic dispersion relations of the form $\omega({\bf k})=\omega(k_x)+\omega(k_y)+\omega(k_z)$, so that the transformed harmonic oscillators will conform a structure that for next neighbors interaction is just a cubic lattice (see this example in Fig. \ref{cubic2D}). 

\begin{figure}[ht]
%\centerline{\includegraphics[width=0.5\textwidth]{chain2.pdf}}
%\centerline{\includegraphics[width=0.5\textwidth]{tresH.pdf}}
%\centerline{\includegraphics[width=0.5\textwidth]{ChainOnly.pdf}}
\centerline{\includegraphics[width=0.45\textwidth]{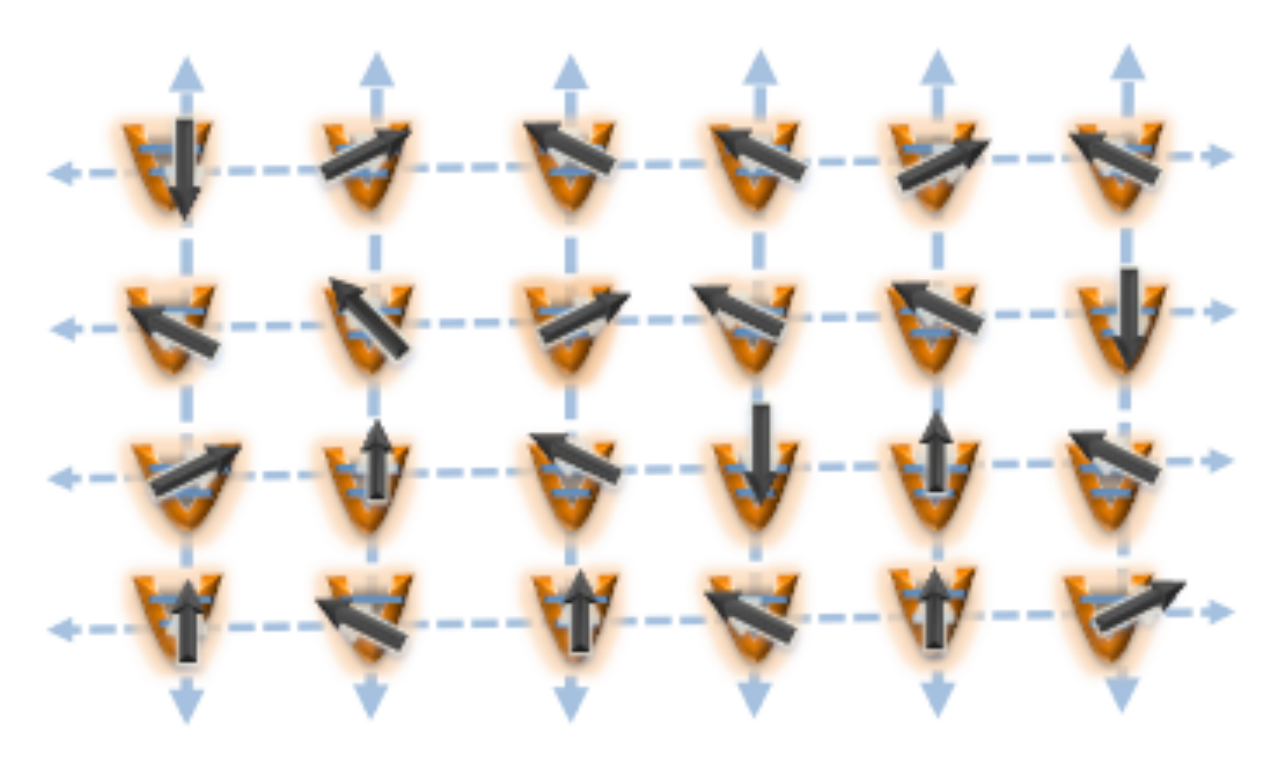}}
\caption{(Color online) A crystal of atoms and transformed modes conforming a cubic structure. It represents a mapping of the form (\ref{fintrans3D}) for an ensemble of atoms arranged in positions ${\bf r}_n={\bf n}h_0$, coupled to the radiation field within a $2D$ photonic crystal, which has dispersion relation $\omega({\bf k})=A+B_x\cos{(k_x h_0)}+B_y\cos{(k_y h_0)}$ \cite{devega2005} (the generalization to three dimensions is straightforward). The representation of the light-matter interaction problem here proposed may be useful to design and analyse schemes to control the flow of light and its absorption. \label{cubic2D}}
\end{figure}

\subsection{Isotropic dispersion relations}

For isotropic dispersion relations, if the atoms are in a one dimensional structure, the problem can still be mapped to a ladder-like structure. Let us write the interaction Hamiltonian (\ref{hEM1}) as
\beqa
H^\ttd_\iint=\frac{g}{\sqrt{M}} \sum_n \sum_q\int d\Omega_{\bf k}\bigg(a_{\bf k} e^{ik_q r_n}+h.c.\bigg)C_n.
\label{hEM1}
\eeqa
where we have assumed that the atoms are placed along the $z$ axis, and that the momentum modulus is discretized with an index $q$. In addition, we consider an isotropy correlation function $\omega({\bf k})=\omega(k)$, and 
$g({\bf k})=g$. 
%The last replacement can be made exact for independent environments, where $\alpha_{nm}^\ttd(t)\approx\delta_{nm}\alpha_{nm}(t)$. In this case, as discussed in \cite{chin2010}, the effects of the environment are integrally encoded in the spectral function $J(\omega)=g^{2}(\epsilon(\omega))\rho_\DDOS(\omega)$, where $\rho_\DDOS(\omega)=\frac{d \epsilon (\omega)}{d\omega}$, and $\epsilon(k)$ is the inverse function of the dispersion $\omega(k)$, i.e. $\epsilon(\omega(x))=x$ \cite{chin2010}.  Thus, different pairs of $g(k)$ and $\omega(k)$ can lead to the same $J(\omega)$, and we can consider $g(k)=g$ in particular, together with a new $\hat{\omega}(k)$ such that the desired spectral density is obtained. 
%In that case, the dynamics of the particles coupled to the harmonic oscillator environment can be described by the equivalent Hamiltonian (\ref{h1EM}), but now with the new $\hat{\omega}(k)$ replacing $\omega(k)$, and $g(k)=1$. 
Also, $g(k)=g$ is a good approximation when $\omega(k)$ gives rise, in the calculus of $\alpha_{nm}^\ttd(t)$, fo a phase that varies much more fast in $k$ than $g(k)$, which can be considered as a constant in comparison \cite{devega2005}. 

Let us consider the transformation $a_{\bf k}=\int d\Omega_{r_p} \sum_{p=0,N-1}U_{\bf p}({\bf k}) b_{\bf r_p}$, with 
\bea
U_{\bf p}({\bf k})= \frac{1}{\sqrt{M}}\sum_{l=0}^\infty \sum_{m=-l}^l i^l Y_{lm}(\theta_p,\phi_p)Y_{l,m}^*(\theta_k,\phi_k)e^{-i k_q r_p},\cr
\eea
where $(r_n,\theta_n,\phi_n)$ and $(k,\theta_k,\phi_k)$ are ${\bf r}_n$ and ${\bf k}$ in spherical coordinates. Also, $Y_{lm}(\theta,\phi)$ are spherical harmonics. With this transformation, the interaction Hamiltonian (\ref{hEM1}) can be written as
\beqa
H^\ttd_\iint&=&\hat{g}\int d\Omega_{r_p} \sum_{p=0,M-1}  \sum_q\int d\Omega_{\bf k}\bigg(\sum_{l=0}^\infty \sum_{m=-l}^l i^l \cr
&\times&Y_{lm}(\theta_p,\phi_p)Y_{l,m}^*(\theta_k,\phi_k)e^{-i k_q (r_p-r_n)}b_{{\bf r}_p}+h.c.\bigg)C_n.\cr
\label{hEM2}
\eeqa
where $\hat{g}=g/M$.
Interestingly, $Y_{00}(\theta,\phi)=\frac{1}{2\sqrt{\pi}}$ for any angle $\theta,\phi$. Thus, inserting $2\sqrt{\pi}Y_{00}(\theta_k,\phi_k)$ in the above expression we can solve the angular integral in $\Omega_k$, 
\bea
\int d\Omega_k Y_{0,0}(\theta_k,\phi_k)Y^*_{l,m}(\theta_k,\phi_k)=\delta_{l0}\delta_{m0},
\eea
with $\int d\Omega_k=\int_0^\pi d\theta\sin(\theta)\int_0^{2\pi} d\phi$, where we have used the property
\bea
\int d\Omega_k Y_{l,m}(\theta_k,\phi_k)Y^*_{l',m'}(\theta_k,\phi_k)=\delta_{ll'}\delta_{mm'}. 
\label{propgen}
\eea
Note that the property 
\bea
\sum_{lm} Y^*_{l,m}(\theta_p,\phi_p)Y_{l,m}(\theta_n,\phi_n)=\delta(\theta_p-\theta_n)\delta(\phi_p-\phi_n).
\eea
is also needed to show that the transformed operators $b_{{\bf r}_p}$ fulfill the proper conmutation relations, so that the transformation is canonical. 
In addition, assuming that $|k|\equiv k_q=2\pi q/(Md_0)$, we find that $ \sum^M_{q=0} e^{i k_q (r_n- r_p)}=M\delta_{np}$. Hence, we find that 
\beqa
H^\ttd_\iint&=&g\sum_n\bigg(B_{0,0,n}+B^\dagger_{0,0,n}\bigg)C_{n},
\label{hEM6sp}
\eeqa
where we have defined $B_{l,m,r_n}=\int d\Omega Y_{lm}(\theta,\phi) b_{r_n,\theta,\phi}$.
Note that because of the property (\ref{propgen}), this operators obey the usual bosonic commutation rules $[B_{l,m,n},B^\dagger_{l',m',n'}]=\delta_{ll'}\delta_{mm'}\delta_{nn'}$.
%
%\bea
%\bigg(\frac{1}{\sqrt{2\pi}}\bigg)^3e^{i{\bf k}\cdot{\bf V}}&=&\sqrt{\frac{2}{\pi}}\sum_{l=0}^\infty \sum_{m=-l}^l i^l j_l(kr_v)Y_{lm}(\theta_v,\phi_v)\cr
%&\times&Y_{l,m}^*(\theta_k,\phi_k), 
%\label{exp2}
%\eea
%
%It is important to assume further that ${\bf r}_n=(r_n,0,0)$, such that 
%\bea
%\bigg(\frac{1}{\sqrt{2\pi}}\bigg)^3e^{i{\bf k}\cdot{\bf r}_n}&=&\sqrt{\frac{2}{\pi}}\sum_{l=0}^\infty \sum_{m=-l}^l i^l j_l(kr_n)Y_{lm}(0,0)\cr
%&\times&Y_{l,m}^*(\theta_k,\phi_k).2
%\label{exp2}
%\eea
%Replacing (\ref{exp1}) and (\ref{exp2}) in (\ref{hEM1sp}), we find 
%\beqa
%H^\ttd_\iint=\sum^N_{n=1}  \bigg(\frac{g}{r_n}\bigg)\bigg(b_{n,\theta_n=0,\phi_n=0} +b^\dagger_{n,\theta_n=0,\phi_n=0}\bigg)C_n,
%\label{hEM1sp2}
%\eeqa
%where we have used the following properties:
%\bea
%\int d\Omega_k Y_{l,m}(\theta_k,\phi_k)Y^*_{l',m'}(\theta_k,\phi_k)=\delta_{ll'}\delta_{mm'}, 
%\eea
%
%and 

Let us now consider $H_B=\sum_{\bf k}\omega({\bf k})a^\dagger_{\bf k}a_{\bf k}$, and transform it as before
%and consider the proposed transformation
\beqa
H_B&=&\sum_{n,p,q}\int d\Omega_k\omega(k_q)\int d\Omega_p\int d\Omega_n U^*_{\bf n}({\bf k})U_{\bf p}({\bf k}) b^\dagger_{\bf r_n} \cr
&\times&b_{\bf r_p}.
\label{hBsp}
\eeqa
In detail, it can be written as 
\beqa
H_B&=&\frac{1}{M}\sum_{n,p,q} \omega(k_q)\int d\Omega_p\int d\Omega_n \sum_{l,l'} \sum_{m,m'} (-i)^l(i^{l'})\cr
&\times& Y^*_{lm}(\theta_n,\phi_n)Y_{l'm'}(\theta_p,\phi_p)F_{ll'mm'}\cr
&\times&e^{i k_q (r_n-r_p)}b^\dagger_{\bf r_n} b_{\bf r_p}.
\label{hBsp2}
\eeqa
where $F_{ll'mm'}=\int d\Omega_k Y_{l,m}(\theta_k,\phi_k) Y_{l',m'}^*(\theta_k,\phi_k)=\delta_{ll'}\delta_{mm'}$ according to (\ref{propgen}). Hence, we find that 
\beqa
H_B&=&\frac{1}{M}\sum_{l,m}  \sum_{n,p}\sum_q\omega(k_q)e^{i k_q (r_n-r_p)}\int d\Omega_p \cr
&\times& Y^*_{lm}(\theta_n,\phi_n) b^\dagger_{\bf r_n} \int d\Omega_n Y_{lm}(\theta_p,\phi_p)b_{\bf r_p}.
\label{hBsp3a}
\eeqa
Or simply,
\beqa
H_B&=&\sum_{lm}\sum_{np} f_{np} B^{\dagger}_{l,m,r_n}B_{l,m,r_p}
\label{hBsp3}
\eeqa
where 
\bea
f_{np}=\frac{1}{M}\sum_q \omega(k_q) e^{i k_q (r_n-r_p)}.
\label{coefmap}
\eea
Indeed, from (\ref{hEM6sp}) and (\ref{hBsp3}), the only modes involved in the dynamics are the isotropic modes, $B_{0,0,r_n}$ and $B^\dagger_{0,0,r_n}$, with discrete positions $r_n$. This shows that for an isotropic dispersion relation, the full $3D$ problem can be mapped into a ladder-like structure of the form (\ref{fintrans}) as in the $1D$ case.

\subsection{Limit of independent environments}
\label{indep}
There is a limit in which the different atoms within the structure evolve as if each of them were interacting with its own environment. This limit is defined when the correlation function decays very fast with the inter-particle distance, such that $\alpha_{nm}(t)\approx\delta_{nm}\alpha_{nm}(t)$. 
Note that this case is one of the most common in quantum optics, and corresponds to atoms emitting independently to each others. In this particular case, the atoms can be assumed to be arranged in any spatial structure and not necessarily a cubic lattice. 

An alternative way to find this limit, can be based on the mapped structure. Indeed, the evolution time scale of the OQS can be estimated as $T_\ddiss\approx 1/\Gamma$, where $\Gamma\approx\Rre[\int_0^\infty ds\alpha_{nn}(s)]$. Then, the distance that an excitation go along one direction through the chain during that time is $L_\ddiss\approx vT_\ddiss$, where $v=d_0f$ is the velocity of the excitation within the chain, with $f$ the average hopping rate between sites. Hence, in the limit where $L_\ddiss\ll d_0P$, or $f/\Gamma\ll P$, the excitation will not have time to travel to adjacent atoms within the structure during the dissipation time, and therefore each atom within the ensemble will be coupled to its own environment. 

In the above discussed limit, the effects of the environment are integrally encoded in the spectral function $J(\omega)=g^{2}(k(\omega))\rho_\DDOS(\omega)$, where $\rho_\DDOS(\omega)=A(k(\omega))=|\frac{d \omega(k)}{dk}|^{-1}_{k=k(\omega)}$, as $A(k)=|\frac{d \omega(k)}{dk}|^{-1}$  \cite{chin2010}.  Thus, different pairs of $g(k)$ and $\omega(k)$ can lead to the same $J(\omega)$, and we can consider $g(k)=g$ in particular, together with a new $\hat{\omega}(k)$ such that $|\frac{d \hat{\omega}(k)}{dk}|^{-1}_{k=k(\omega)}=J(\omega)$, where now $k(\omega)$ is the inverse of $\hat{\omega}(k)$ (See Appendix B for an example).

\section{Link to coupled cavity QED}
When considering $H_S=\sum_n\omega_n\sigma_n^\dagger\sigma_n$, the Hamiltonian (\ref{fintrans}) has the form of that of $M$ cavities with a resonant mode $b_n$ and energy $f_n=f_{nn}$, each cavity containing a single atom of frequency $\omega_n$ coupled to the cavity mode with strength $g$. Then,  the second term of (\ref{fintrans}) for $f_{nm}$ with $n\neq m$, can be seen as a coupling between the different cavities, where each cavity site is described by the well-known Jaynes-Cummings model. 
%In this model, the atom-photon interaction gives rise to eigen-energies that at a site $n$ have the form $E_{|\pm,j\rangle}=jf_{n}\pm\xi(j)-\Delta/2$, where $\Delta=\omega_n-f_{n}$, and $\xi(j)=\sqrt{jg^2+\Delta^2/4}$ being the $j$ photon generalized Rabi frequency. The increasing energy separation with $j$ reflects the strong non-linearities appearing in the system due to the atom-photon interaction. 
The non-linearities in this system lead to an on-site repulsion, which combined with the hopping term between the cavity modes $b_n$ leads to a Bose-Hubbard-like dynamics where quantum phase transitions of light can be described in the ground state between a Mott-like phase (i.e. the photon blockade regime) and a superfluid phase
%. For a coupled cavity system where each cavity contains an atom,  this Bose-Hubbard dynamics leads to a quantum phase transition of light induced by the coupling with the atoms 
\cite{greentree2006,hartmann2006,angelakis2007,schiro2012}. 
%Indeed, in the limit of weak copling between cavities, a photon blockade regime occurs were photons do not flow anymore between cavities. 
In addition, strong signatures of photon blockade have also been observed in the non-equilibrium dynamics of weakly coupled cavity arrays \cite{nissen2012}. 
%This suggests an interesting parallelism between the photon-blockade regime and a regime with strong non-Markovianity, as it will be discussed in the example that follows. 
Thus, in the transformed system, we shall expect similar ground state and dynamics than in coupled cavity arrays. The main difference is that in the transformed system, the atoms may be directly coupled with each others through resonant dipole-dipole interactions, while in the coupled cavity case this might be more difficult to achieve. Also, as discussed above, for some particular environments the modes may be connected beyond next neighbors. 
%\textit{Evolution in structured environment--}
%
%An analysis of the exact evolution equations of an OQS \cite{alonso2007} leads to the conclusion that the only quantity necessary to describe the coupling with the environment is the so-called correlation function,
%\beqa
%\alpha^{1D}_{nm}(t)=\int^{k_\mmax}_{-k_\mmax} dk g^2(k) e^{ikr_{nm}-i\omega(k)t}.
%\label{h1corr}
%\eeqa
%Here $r_{nm}=r_n-r_m=d_0(n-m)$, where $d_0$ is the mean inter-particle spacing. $\alpha_{nm}(t)=u_nu_m\int^{k_s}_{-k_s} dk \phi^*_n(k)\phi_m(k)e^{-i\hat{\omega}(k)t}$ and (\ref{h1corr}) is that the earlier does not depend on any function $g(k)$. In many cases, this problem can be overcome by approximating $g(k)\approx g(k_0)$, with $k_0$ the resonant wavelength, when it varies much less in $k$ than the exponential phase \cite{devega2005}.

For the case of two dimensional environments, the transformed system may also give rise to a coupled cavity array of the form of Fig. (\ref{cubic2D}) provided that the dispersion relation can be written as $\omega({\bf k})=\omega(k_x)+\omega(k_y)$. Then, the coupling $f_{{\bf n m}}=f_{n_x,m_x}\delta_{n_y,m_y}+f_{n_y,m_y}\delta_{n_x,m_x}$, i.e. it will only connect sites within the $x$ or $y$ directions, but never across the diagonal. Note that a similar reasoning can be followed for the case of three dimensional environments. 

\section{Application: Atoms within a 1D photonic crystal}
We shall consider the dynamics of atoms or quantum dots embedded in a $1D$ photonic crystal structure \cite{yablonovitch1987,john1987}, considering that the other two directions are non-dispersive. This problem can be described with the Hamiltonian (\ref{h1}), considering a dispersion relation of the form $\omega(k)=A+B\cos((k-k_0)h_0)$ \cite{devega2005}, where $k_0=\pi/h_0$, $h_0$ is the linear size of the unit cell of a cubic lattice, and $k$ runs from $-k_0$ to $k_0$ within the first Brillouin zone (see Fig. (\ref{dispersion})).
% Actualizar:
PCs typically consist of a low-dielectric-constant network, inserted in a high-dielectric-constant backbone in a periodic structure. There is a large variety of photonic crystals giving rise to a complete one, two or three dimensional band gap \cite{pcbook}. 
%Some examples are the inverse diamond structure formed by overlapping air spheres in a dielectric material \cite{HCS90}, the woodpile structure (a variation of the diamond structure) \cite{L98}, $SiO_{2}$ spheres in a $InP$ backbone \cite{M99}, or spiral shaped rods arranged in simple cubic (SC), face-centered cubic (FCC) on any body centered cubic (BCC) lattices \cite{CN98}. For an extended  review of the current state of photonic band structure theory, both experiments and applications, see \cite{KB02,WJ03}. 
Of special interest is the analysis and control of the spontaneous emission, which can be dramatically modified by the presence of such band-gap dispersion relation \cite{john1994,florescu2001,devega2005,giraldi2011}. This may have important applications ranging from miniature lasers and light-emitting diodes, to single-photon sources for quantum information, and to solar energy harvesting \cite{lodahl2004}. Particularly, experimental progress in the control of spontaneous emission by manipulating optical cavity modes and quantum dots within photonic crystals have demonstrated that the spontaneous emission from light emitters embedded in photonic crystals can be suppressed by the so-called photonic bandgap, whereas the emission efficiency in the direction where optical modes exist can be enhanced \cite{noda2007,thompson2013}. Partial dissipation is also observed in Fano-Anderson-like models, where one or more quantum emitters couples to a finite band of modes \cite{gaveau1995}. Recent proposals \cite{hung2013,goban2013} explore the atom-atom interactions that may be produced in these materials, and which are mediated by a strong light-matter interaction. Hence, being able to analyze such effects with analytic tools that allow to explore regimes beyond the Markov and weak coupling approximations may be of extreme importance to understand experiments and lead to further developments. 
 
%Of special interest for optical applications are the PCs with the photonic gap in the IR and visible regimes. One of the most recently developed PC with an optical PBG, consists in silica spheres, with diameters between $600$ and $1000 \ nm.$, periodically inserted in bulk silicon \cite{BCG00}. The size and disposition of the spheres gives rise to the formation of a $5\%$ complete 3D PBG centered near 1.5 $\mu m$, which has been experimentally observed. 

\begin{figure}[ht]
%\centerline{\includegraphics[width=0.5\textwidth]{Figure12_3.pdf}}
\centerline{\includegraphics[width=0.4\textwidth]{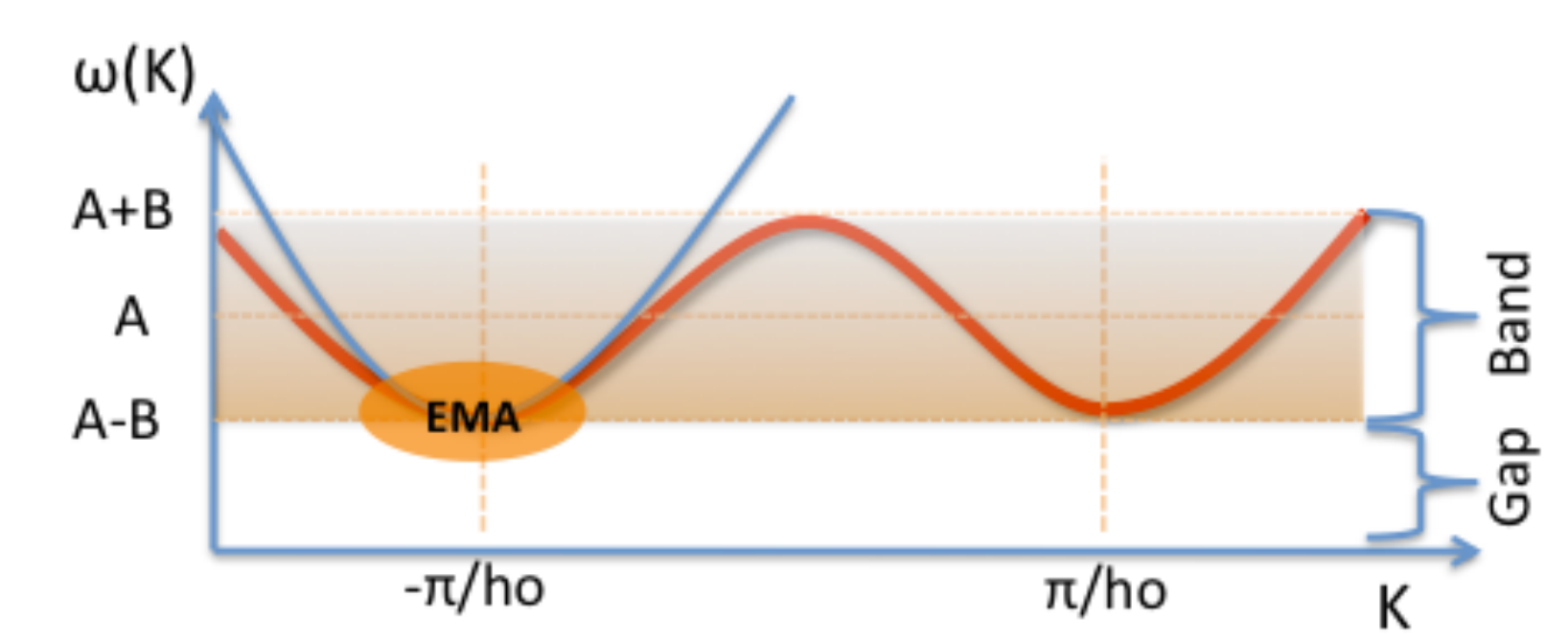}}
\caption{(Color online) Dispersion relation of the radiation field within a photonic crystal. It shows the relationship between the frequency of the photons and the different values of the wave vector within the first Brillouin zone (ranging from $[-\pi/h_0,\pi/h_0]$). The dispersion relation is not defined for frequency values within the range from $0$ and $A-B$, and larger than $A+B$. These values correspond to a gap in the photonic density of states. The orange oval denotes to the approximate parameter region where $\delta/B\ll 1$, so that an effective mass approximation (EMA) can be considered (see discussion). \label{dispersion}}
\end{figure}

The correlation function corresponding to (\ref{h1}) can be written as (\ref{correl1D}) or, considering that the all atomic frequencies are $\omega_0$,
\begin{eqnarray}
\alpha_{nm}(\tau)&=&\hat{\gamma}(\frac{h_0}{2\pi})\sum_{\sigma}\int_{1BZ}dk\frac{ |\hat{e}_{{\bf k},\sigma}\cdot \hat{u}_d|^{2}}{\omega(k)}e^{ikr_{nm}-i\omega(k)\tau}\cr
&=&\gamma\int_{-\frac{\pi}{h_0}}^{\frac{\pi}{h_0}}dk e^{ikr_{nm}-i\omega(k)\tau},
\label{generalcorr5}
\end{eqnarray}  
where we have used an approximation similar to (\ref{corr3D}), and defined again $\gamma=\hat{\gamma}(\frac{h_0}{2 \pi}) \sum_{\sigma}\frac{ |\hat{e}_{k_\eeff,\sigma}\cdot \hat{u}_d|^{2}}{\omega(k_\eeff)}$. Here, $\alpha_{nn}(t)=\gamma e^{-iAt}J_0(Bt)$, with $J_0(t)$ the zeroth order Bessel function. 

%We will consider $A= 10^{15}$ Hz as the unit frequency of the problem.
%$\tau_c=1/B$ is the relaxation time scale of the environment

%The decaying of such function is given by $\tau_c=1/B$, and gives the relaxation time scale of the environment. 

%a fucntion that decays with a correlation time   In fact, the system relaxation is approximately given by $1/g$, while $\tau_c=1/B$ is the relaxation time scale of the environment, i.e. the time in which the correlation function,  $\alpha_{nn}(t)=e^{-iAt}J_0(Bt)$ (with $J_0(t)$ the zeroth order Bessel function and chosing $n=m$), decays. Hence, the photon blockade regime, $B/g\ll 1$,  indeed corresponds, in the language of OQS, to a strong coupling regime. 

The transformed Hamiltonian takes the form (\ref{fintrans}), with
$f_{nn}=\sum_{q=0}^{M-1} e^{-ik_qr_n}e^{ik_qr_n} \omega(k)=A$, and $f_{n,m}=\sum_{q=0}^{M-1} e^{-ik_qr_n}e^{ik_qr_m}\omega(k)=\delta_{m,n\pm 1}\frac{B}{2}$. 
The resulting Hamiltonian is 
\bea
H_{\textit PC}&=&H_S+A\sum_n b_n^\dagger b_n+\frac{B}{2}\sum_{n=1,\cdots M}\bigg(b_n^\dagger b_{n+1}\cr
&+&b_{n+1}^\dagger b_{n}\bigg)+g\sum_{n=1,\cdots M}  C_n (b_n+b_n^\dagger),
\label{mappedPC}
\eea
with $C_n=\sigma_n+\sigma_n^+$, and 
\bea
H_S=\sum_{j}\omega_0\sigma_j^+\sigma_j-\sum_{\langle jl\rangle}J\sigma_j^+\sigma_l.
\eea
For the sake of simplicity in the calculations, we shall consider the rotating wave approximation in the former Hamiltonian, so that the terms $b_n^\dagger \sigma^+$ and $b_n \sigma_n$, that simultaneously creates (and annihilates) one photon and one excitation in the atomic lattice are discarded. 
%Here we have a ladder system where a chain of spins or oscillators forming the many body OQS, and represented by the creation (annihilation) operators $a^\dagger_n (a_n)$, are coupled to a chain of oscillators corresponding to the environment with operators $b^\dagger_n (b_n)$, through a coupling operator of the form $C_n$. Such coupling operator should be an Hermitian conbination of OQS operators, for instance $C_n=a_n+a_n^\dagger$.
% UNITS
\subsection{Units of the problem}
\label{units}
A brief comment is here in order regarding the units of the problem. The parameters that characterize impurities in a photonic crystal with a gap in the optical region \cite{florescu2001,devega2005} are $\omega_{0}\sim 10^{15}$ Hz, $d_{21}\sim 10^{-29}$Cm ($10^{-28}$Cm for quantum dots), $h_0\sim 10^{-6}-10^{-7}$m, and $A\sim 10^{15}$Hz. With these values, a realistic strength of the coupling constant appearing in the Hamiltonian is around $g\approx g_{n}(k_0)=-i \sqrt{\frac{1}{2\hbar \omega(k) \epsilon_0\nu}}\omega_0d_{12}$ of the order of GHz-THz. Note that this quantity shall not be confused with the decaying rate, which is given by $\Gamma\approx \Rre[\int_0^\infty ds\alpha_{nn}(s)]$, and therefore is related to the constant $\gamma$ in (\ref{generalcorr5}). 
The correlation function, when appearing in the evolution equations in interaction image with respect to the system Hamiltonian can be re-written as $\alpha^\iint_{nm}(t)= \gamma\int_{-\frac{\pi}{h_0}}^{\frac{\pi}{h_0}}dk e^{ikr_{nm}+i(\Delta-B\cos((k-k_0)h_0)t)}$, where $\Delta=\omega_0-A$. Hence, the atomic dynamics will only depend on the different values of $\Delta$, $B$ and $g$. Here, we will chose an energy scale $\xi$, and assume that all the quantities are re-normalized in this scale, i.e. $\tilde{\Delta}=\delta/\xi$, $\tilde{J}=J/\xi$, $\tilde{B}=B/\xi$, $\tilde{g}=g/\xi$, and $\tilde{t}=t\xi$, although in the following the tilde is omitted for simplicity in the notation. The choice $\xi$ between the range of GHz-THz will lead us to typical values for photonic crystal coupling strengths, what for the chosen parameters will force the band width $B$ to be within a similar range. The choice of other scaling parameters may correspond to other applications different from photonic crystals, like for instance atoms in optical lattices \cite{devega2008,navarrete2010}. 

In the limit where $(k-k_0)h_0\ll 1$ (i.e $k\ll 2/ h_0$), we may expand the dispersion relation $\omega(k)=A+B\cos(k-k_0)$ as $\omega_k\approx\omega_c+B/2(k-k_0)^2$ \cite{devega2005}. This last expression for the dispersion relation corresponds to the one obtained with the so-called effective mass approximation \cite{john1994,florescu2001}, and correspond in frequencies to the choice $\delta/B\ll 1$ (See Fig.(\ref{dispersion})).   
Hence, in the effective mass approximation, the correlation function of the environment appearing in the atomic evolution equations in interaction image with respect to the system, is written as $\alpha^\iint_{nm}(t)\approx \gamma\int_{-\frac{\pi}{h_0}}^{\frac{\pi}{h_0}}dk e^{ikr_{nm}-i(\delta+B/2(k-k_0)^2)t}$, where $\delta=\omega_0-\omega_c$. In this case, the atomic dynamics only depend on the different values of $\delta$ (see also \cite{john1994,florescu2001} for more details), the band width $B$, and the coupling strength $g$. 

%Let us now chose an energy scale $\xi$, and writte all the quantities renormalized in this scale, $\tilde{\delta}=\delta/\xi$, $\tilde{B}=B/\xi$, and $\tilde{g}=g/\xi$. Hence, all curves such that $\delta/B\ll 1$, correspond to physically realistic coupling strengths as long as we consider $B\sim 10^7$, so that the general scale of the problem is fixed as $\xi=10^7$.  In the following, all plots are given in units of $\xi$, but we have omitted for simplicity the tilde in the quantities $A$, $B$, $\omega_0$ and $g$. We note in addition that some of the plots discussed here go beyond the case $\delta/B\ll 1$, so that they might not correspond to the most usual parameters found in photonic crystals. However, these curves have been included for completeness of the analysis, and in view of other applications different from photonic crystals, like for instance atoms in optical lattices \cite{devega2008,navarrete2010}. 

\subsection{Comparison to the master equation}

Let us consider the solution of this problem according to a master equation. This equation describes the evolution of the reduced density operator of the atoms, which is obtained by tracing out all the environment degrees of freedom as $\rho_s(t)=\Ttr_B[\rho_\ttot(t)]$. Up to second order in the coupling parameter between system and environment, $g$, the master equation corresponding to (\ref{h1}) is given by \cite{breuerbook}
\begin{eqnarray}
\frac{d\rho_s  (t)}{dt}&=&-i[H_S (t),\rho_s (t) ]\cr
%&+&\int_0^t d\tau \sum_{lj} \alpha_{lj}^{+*}(t-\tau) [L_l^\dagger ,\rho_s (t)  V_{\tau-t}L_j]\nonumber\\
%&+&\int_0^t d\tau \sum_{lj} \alpha^{+}_{lj}(t-\tau) [V_{\tau-t } L_j^\dagger \rho_s (t) ,L_l]\nonumber\\
&+&\int_0^t d\tau \sum_{lj} \alpha_{lj}(t-\tau)[L_j(\tau-t)\rho_s (t) ,L_l^{\dagger}]\nonumber\\
&+&\int_0^t d\tau  \sum_{lj} \alpha^{*}_{lj} (t-\tau)[L_l,\rho_s (t) L_j^{\dagger}(\tau-t)].
\label{icc20}
\end{eqnarray}
with $L_j(t)=e^{iH_S t}L_je^{-iH_S t}$, and $\alpha_{lj} (t-\tau)= g^2\sum_k e^{i k r_{lj}-i\omega_{\bf k} (t-\tau)}$, where $r_{lj}=d_0(l-j)$. To derive this equation, the so-called Born approximation has been assumed. Hence, the correlations between the system and the environment have been neglected, so that $\rho_\ttot(t)=\rho_s(t)\otimes\rho_B$, where $\rho_B$ is the environment density operator considered always in its equilibrium state.

%%% COMPARISON WITH MASTER EQUATION

We now compare in Fig. (\ref{chainPC0}) the population dynamics of $N$ atoms given by the master equation (\ref{icc20}), and by a Schr{\"o}dinger equation for the mapped Hamiltonian $H_{\textit PC}$. 
This simple example shows the power of the proposed scheme. While the master equation clearly fails after a few time steps giving rise, in some cases, to non-physical results (having a reduced density matrix $\rho_S$ with negative eigenvalues), the Schr{\"o}dinger equation for $H_{\textit PC}$ gives the correct evolution. The number of oscillators needed is of the order of the time scale to be reached, i.e. $M=100$ oscillators.

 \begin{figure}[ht]
%\centerline{\includegraphics[width=0.43\textwidth]{Mix1a.pdf}}
%\centerline{\includegraphics[width=0.45\textwidth]{compara1.pdf}}
\centerline{\includegraphics[width=0.45\textwidth]{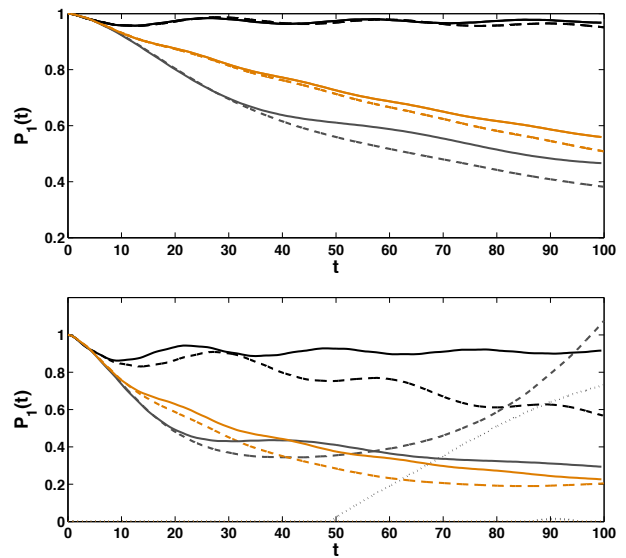}}
\caption{(Color online) Evolution of the population of two atoms (black and red curves respectively), for $B=0.5$, $A=1$, and $g=0.05$ (upper plot) and $g=0.1$ (lower plot), considering  $\omega_0=0.3$, $J=0.$, and $|\psi_0\rangle=|1\rangle|0\rangle$. Solid and dashed curves correspond respectively to the result obtained using $H_{\textit PC}$ or a master equation up to $g^2$ \cite{breuerbook} derived for (\ref{h1}). Black, grey (with divergent dashed curve in the lower panel) and orange curves correspond to $B=0.5,0.8,1$ respectively. Dotted lines in the plot below represent the sum of negative eigenvalues of the reduced density operator $\rho_S(t)$ obtained from the master equation. For $B=0.8$ the quantity grows very large, indicating the inaccuracy of the result. See units in Sect. (\ref{units}). \label{chainPC0}}
\end{figure} 

%%%% DECAY WITH N FOR J=0.1
We have seen that this model shows a very rich dynamics that the mapping unveils well beyond the weak coupling approximation. In addition, it provides a tool to analyze the formation of cavities within photonic crystal structures, particularly within the gap region. As it can already be seen in Fig (\ref{chainPC0}), for certain parameter regimes the atomic population does not fully decay, which is connected to the fact that the atom is interacting with just a few environment modes. 
%Thus, the mapping provides an accurate description of PC-cavities. 
The mapping shows directly that these few environment modes are just the ones within the chain which are located in the neighborhood of the atom or emitter. We will discuss these ideas in the following for two different cases, $N=2$ and $N=15$ atoms and a single excitation in the system. %The results have been qualitatively confirmed for $2$ excitation (not shown here).

%This as it can be seen in Fig. (\ref{chainPC0}), for $N=2$ and $N=10$ and considering different values of the hopping between the atoms . 

\subsection{Formation of cavities}
The appearance of a non-zero steady state population depends highly on whether the atomic frequency is within the band or the gap (see Fig. (\ref{dispersion})). This can be seen in Fig. (\ref{chain3D}), that presents a density plot of the time average of the population $P_T(t)=\frac{1}{t}\sum_{j=1,N}\int_0^t ds\langle\sigma_j^+(t)\sigma_j(t)\rangle$ at $t=300$ with respect to $B$ and $\omega_0$. When varying the hopping rate $B$, a crossover is observed between the regime where the atomic population does not vanish in the long time limit, and a regime where full relaxation is observed. Indeed, a similar crossover was observed in the region of small $\delta=\omega_0-\omega_c$ \cite{john1994}, %where $\omega_c$ is the band-gap edge \cite{john1994}, 
when analyzing the problem within the effective mass approximation.  Also, as seeing in Fig. (\ref{chain3D}), at longer times the contrast between the lighter regions (with non-vanishing atomic population), and the darker regions (with vanishing atomic population) is expected to become stronger.
In other words, the black region in the upper panel corresponds to parameters where the atomic population is slowly decaying to zero. To see this, the lower panel shows the time evolution of $P_T(t)$ at even longer times ($t=800$) than the time at which the upper density plot is represented. Such time evolution is displayed for the parameters corresponding to the points marked in circles in the upper figure.  It is observed that for $B=0.6$ the values of $P_T(t)$ at $t=800$ (of the order of $0.1$ and $0.5$) are smaller than the ones for $t=300$ ($0.2$ and $0.1$ respectively), showing a slow decaying. 

\begin{figure}[ht]
%\centerline{\includegraphics[width=0.5\textwidth]{Figure12_3.pdf}}
%\centerline{\includegraphics[width=0.5\textwidth]{Multi3D.pdf}}
%\centerline{\includegraphics[width=0.5\textwidth]{New3Db2.pdf}}
\centerline{\includegraphics[width=0.45\textwidth]{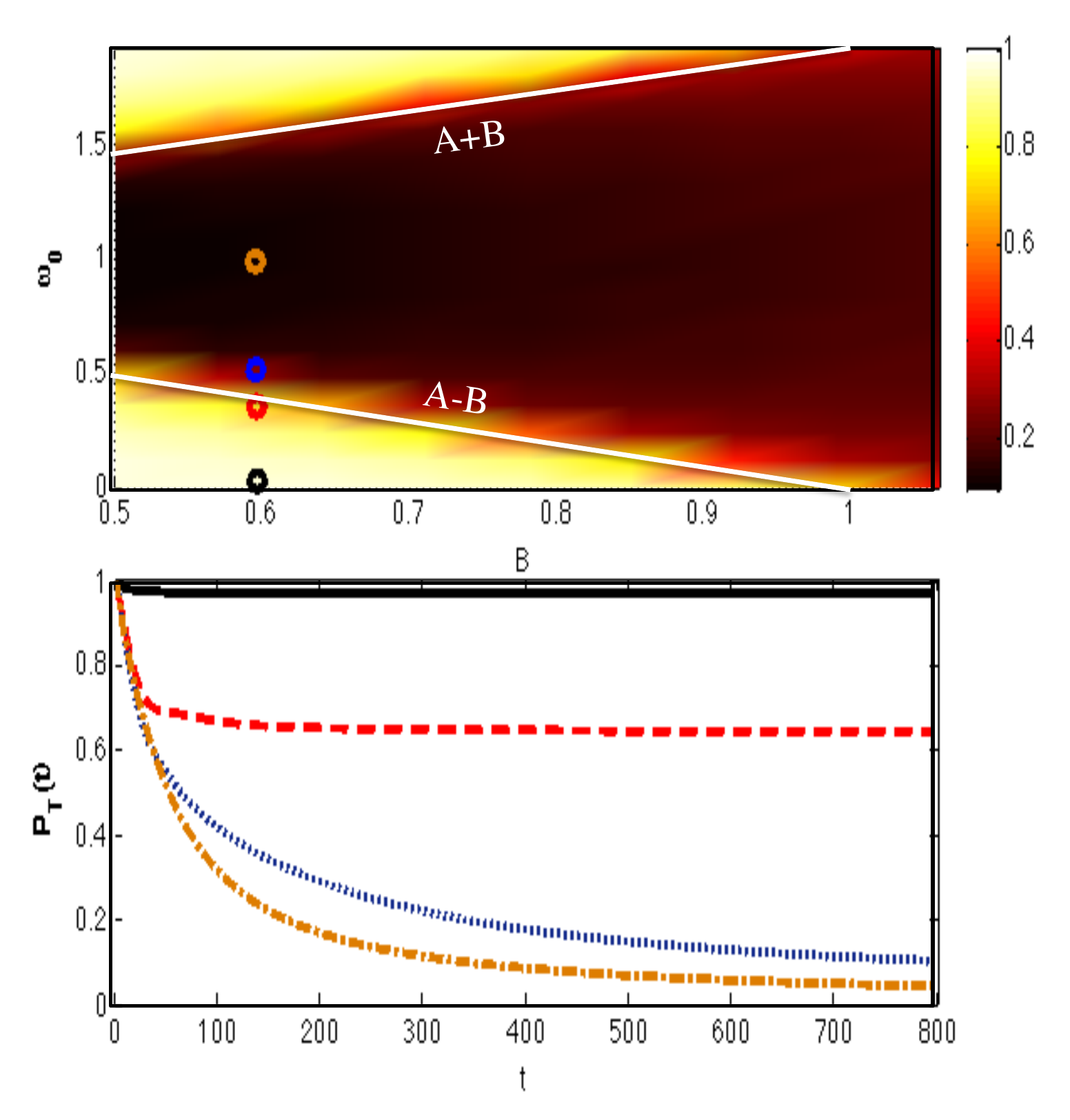}}
\caption{(Color online) Upper plot: Density plot of the time average of the total population $P_T(t)$ at $t=300$ with $\omega_0$ and $B$, with $A=1.$, $g=0.1$ and $J=0$. The lower panel represents the time evolution of $P_T(t)$ at even longer times ($t=800$) for the parameters corresponding to the points marked in the upper figure, i.e. $B=0.6$ and $\omega_0=0,0.4,0.5,1$ (solid black, dashed red, dotted blue and dot-dashed orange respectively).  See units in Sect. (\ref{units}).
\label{chain3D}}
\end{figure}

%We take such time average because the atomic population is a highly oscillating quantity. 

%%% POLARITONS
%In dissipative systems coupled to an environment at zero temperature, one may expect in principle a complete relaxation of the atomic system to its ground state. 
The incomplete relaxation occurring for atomic frequencies within the gap occurs because of the presence of polaritons (known as photon-atom bound states in the photonic crystal literature \cite{john1994,devega2005}). Due to these polaritons, part of the energy initially in the atomic system stays trapped nearby the atoms, and therefore it is never irreversibly lost. The presence of these polaritons, which represent highly correlated atom-photon states, explains partially the failure of the master equation (\ref{icc20}), that as mentioned above is based on the Born approximation and thus neglect any system-environment correlation. Atom-photon bound states are known to lead to non-Markovian dynamics \cite{tong2010}, and such non-Markovianity can be even qualified by the recently derived non-Markovianity measures \cite{breuer2009,rivas2009,rivas2014}. This is done for instance in \cite{tufarelli2014} for the case of an atom radiating in a one-dimensional photonic crystal waveguide in the presence of a mirror. However, it is important to note that even if such bound states are not formed, non-Markovianity can still be significant. This occurs for instance in an atom coupled to a cavity mode, a system where bound states are not formed but which still presents a strong non-Markovianity  \cite{laine2010} (see also \cite{lambropoulos2000}). In our case, at the parameter regimes where the system has no atom-photon bound states, the excitations will flow away through the harmonic oscillator chain so that the OQS will eventually relax to its ground state. However, as noted above some back-flow of information (leading to a temporal increase of the OQS’ coherences) may occur at the environment correlation time scales, and thus non-Markovian effects may still be present.

%%% POLARITONES

The mapped Hamiltonian (\ref{mappedPC}) allows to see clearly the presence of such polaritons in the system. Considering a single excitation, the system basis can be written as a set of atomic-type of states ($|\psi^\aat_1\rangle=|1,0\rangle|vac\rangle$ and $|\psi^\aat_2\rangle=|0,1\rangle|vac\rangle$ for $N=2)$, where the excitation is in the atomic degrees of freedom, and environment type of states ($|\psi^\eenv_j\rangle=b_j^\dagger|0,0,\cdots,0\rangle$, for $j=N+1,\cdots,M$), where the excitation is contained within one of the chain modes. An exact diagonalization of the Hamiltonian in such basis gives rise to four eigenvectors $|P_j\rangle$ that are combination of states $|\psi^\aat_n\rangle$ where the excitation is in the atoms, and states $|\psi^\eenv_j\rangle$ where the excitation is in near-neighbor modes of the atoms. Here we consider these eigenvectors as polaritons, in the sense that they are conformed by an atomic part and a photonic part. In addition, because they are both eigenstates of $H$ and have no overlap with sites beyond near-neighbors, the subspace they span is what is known in the literature as an invariant subspace \cite{caruso2009}. In such subspaces the excitations are trapped and never flow away to other states external to the subspace. Naturally, if the system is initially prepared in one of the invariant states, then there will be no dynamics. Similarly, the fraction of the initial state corresponding to this invariant states will not vary with the dynamics. Indeed, the initial state $|\Psi_0\rangle=|\psi_1^\aat\rangle$ (corresponding to the first atom initially excited) can be proyected in the basis of polaritonic $|P_j\rangle$ and non-polaritonic or non-invariant eigenvectors $|\phi_j\rangle$, what leads to  $|\Psi_0\rangle=\sum_j a_j |P_j\rangle+\sum_j b_j|\phi_j\rangle$, where $a_j=\langle P_j|\Psi_0\rangle$ and $b_j=\langle\phi_j|\Psi_0\rangle$. Then, because of the non-degeneracy of the Hamiltonian spectra, the amount of population trapped in the invariant subspace can be calculated just as $P_{\ppol}=\sum_j |a_j|^2$. Since polaritons combine atomic and photonic degrees of freedom, this quantity is an upper bound to the final state population within the atomic system. 

Fig. (\ref{polar1}) represents $P_{\ppol}$ with zero hopping rate (upper plot), and finite hopping rate (lower plot). The upper panel presents a very similar profile to the upper panel of (\ref{chain3D}), which reflects the atomic population at long times ($t=300$). The only difference is that in the upper plot of (\ref{polar1}), the population within the band (i.e. for frequencies $A-B< \omega_0< A+B$) vanishes completely, whereas in the plot representing the atomic population the population has not jet decayed completely at $t=300$. A further difference appears to be the fact that the polariton analysis predicts a smaller steady state population, of the order of $0.5$, in the upper gap (i.e. for frequencies above $A+B$) than in the lower gap (i.e. for frequencies below $A-B$), whereas the upper panel of (\ref{chain3D}) shows an equal solution for both gaps. Hence, it can be concluded that the upper gap suffers some decaying, that is nevertheless very slow and cannot be captured at the time scale in which (\ref{chain3D}) is plotted. 

The lower panel in in Fig. (\ref{polar1}) gives the polariton population when considering $J=0.25$. Some of the features within this plot can be qualitatively explained. For instance, the fact that the black region is displaced with respect to the black region in the upper panel, can be explained because the eigen-energies of the new system Hamiltonian, $\hat{H}_S=(\omega_S-\Delta_\llamb)(|0,1\rangle\langle 0,1|+|1,0\rangle\langle 1,0|)+\omega_g|0,0\rangle\langle 0,0|-J\sum_j (|1,0\rangle\langle 0,1|+|0,1\rangle\langle 1,0|)$, and therefore the relevant energy transitions are no longer $\omega_0$ and $0$ (or $\omega_s$ and $\omega_g$ without a re-normalization such that $\omega_0=\omega_s-\omega_g$). In the effective system Hamiltonian, the action of the environment can be included approximately as a Lamb shift, $\Delta_\llamb={\mathcal Im}[\Gamma_0]$, with $\Gamma_0=\int_0^\infty d\tau\alpha_{jj}(\tau)$ the so-called dissipation rate, and $\alpha_{jl}(t)$ given by (\ref{generalcorr5}). Indeed, when diagonalizing $\hat{H}_S$, we find the eigenvectors $|\phi_0\rangle=|0,0\rangle$, $|\phi_1\rangle=\frac{1}{\sqrt{2}}(|0,1\rangle-|1,0\rangle)$, and $|\phi_2\rangle=\frac{1}{\sqrt{2}}(|0,1\rangle+|1,0\rangle)$ corresponding to the eigen-energies $E_0=\omega_g$, $E_1=\omega_S-\Delta_\llamb-J$ and $E_2=\omega_S-\Delta_\llamb+J$ respectively. Hence, two different transitions of the OQS shall be considered, $\Delta_1=E_1-E_0=\omega_0-J$, and $\Delta_2=E_2-E_0=\omega_0+J$ (here we have discarded the Lamb shift, that can be neglected for sufficiently small couplings). When both transitions lie within the lower gap, i.e. $\Delta_1<A-B$, $\Delta_2<A-B$, we are in observe full preservation of the polariton population. For the parameters in Fig. (\ref{polar1}), this corresponds to the case when $\omega_0<0.25$ (white region in the lower panel). However, when $\Delta_1$ is within the gap, but $\Delta_2$ is within the band, i.e. in the region where $A-B-J<\omega_0<A-B+J$, only half of the population is lost, and therefore $P_{\ppol}=0.5$. The quantities at the two sides of the inequality mark the boundaries of the orange region within the lower panel of Fig. (\ref{polar1}), in our case given by $0.25<\omega_0<0.75$. 

In addition, Fig. (\ref{polar2}) shows the amount of population trapped in the form of polariton depending on $J$ and for different values of the atomic frequency. In detail, the curves correspond to values of $\omega_0$ ranging from $0.1$ to $0.5$ (and from $0.6$ to $1$ in the inset) with intervals of $\Delta\omega_0=0.1$. It can be observed that there is a certain value of $J$, which depends on $\omega_0$, up to which the trapped population remains equal to $0.5$. Indeed, following a similar analysis as before, it can be concluded that the full width at half maximum of each curve is given by the point in energy where the transition $\Delta_2$ enters into the band, while $\Delta_1$ remains in the gap. This corresponds for instance to $J=A-B-\omega_0=0.4$ for $\omega_0=0.1$ or $J=0.1$ for $\omega_0=0.4$, and also explains why the different curves are displaced by $\Delta\omega_0$. A similar analysis can be made to analyse the insert within the figure, which shows the permanence of polaritons for atomic frequencies within the band. The proportion of atomic component in the polariton varies for each particular values of $\omega_0$ and $J$. This can be seen from the dotted curves, which represent the total atomic population $P_T(t)$ at $t=500$. For certain values of $J$ this atomic population is still higher than the population within the polariton, which reflects the fact that the population has not completely relaxed to its steady state value (not shown here). 

%Indeed, figure (\ref{polar3}) shows, for different $\omega_0$ that finite values of $J$ gives rise to a slow decay to a steady state value of $0.5$. 

\begin{figure}[ht]
%\centerline{\includegraphics[width=0.5\textwidth]{Figure12_3.pdf}}
%\centerline{\includegraphics[width=0.5\textwidth]{Multi3D.pdf}}
\centerline{\includegraphics[width=0.45\textwidth]{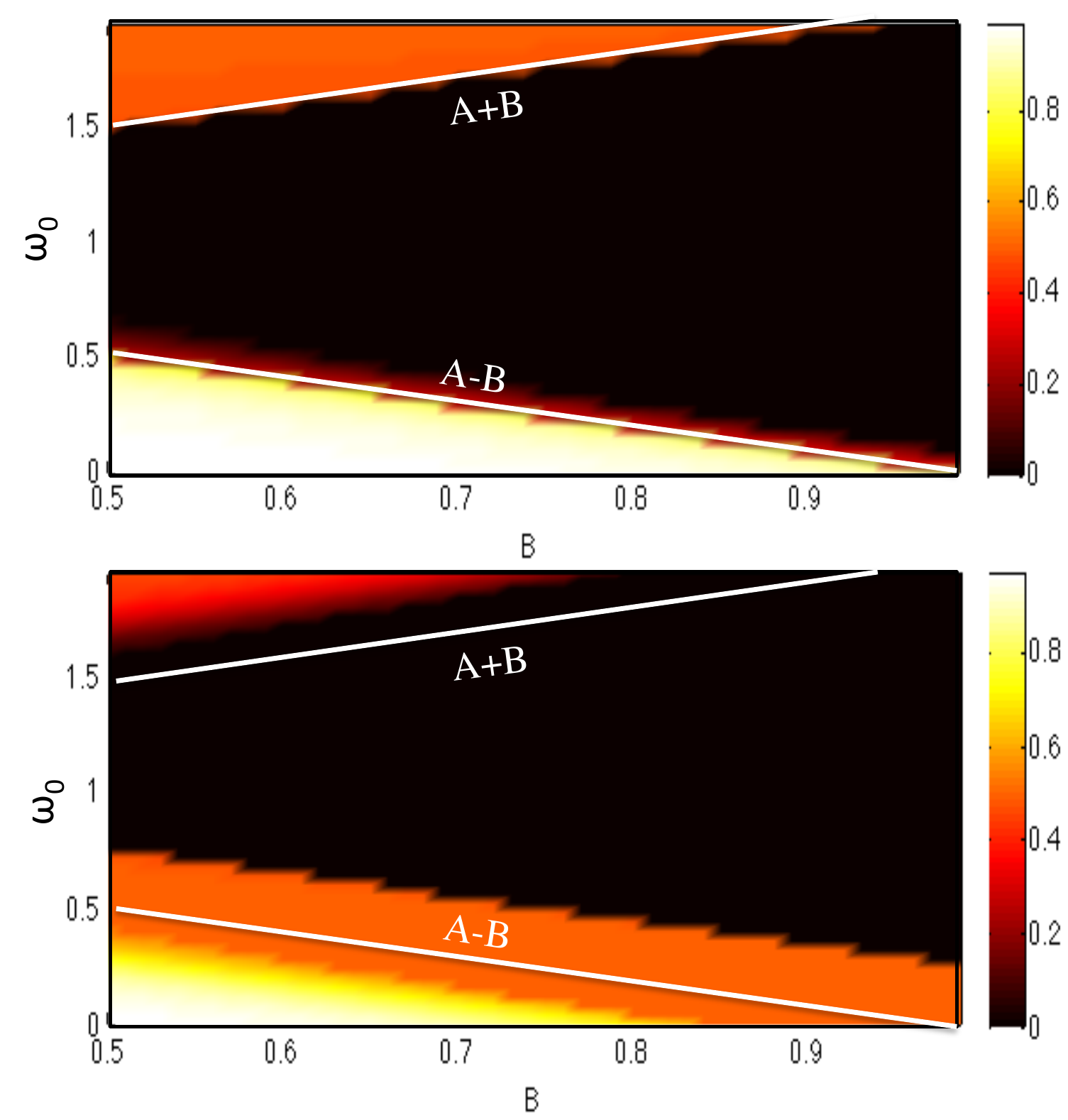}}
\caption{(Color online) Density plot that represents the total population trapped in the polariton state $P_{\ppol}$ with respect to $\omega_0$ and $B$, when considering $g=0.1$, $A=1$, $N=2$ and the initial state $|\Psi_0\rangle=|1\rangle|0\rangle|{\textmd vac}\rangle$. The upper plot corresponds to the case when $J=0$, and the lower plot represents the case $J=0.25$. See units in Sect. (\ref{units}).
\label{polar1}}
\end{figure}

% Figura de los polaritones frente a J
%Fig. (\ref{chainPC2}) represents the values of $P_T(t)$ at $t=500$ for different values of $J$, and considering different values of the atomic frequency. It can be seen that when $J$ is large enough the atomic population is efficiently preserved independently of the value of the atomic frequency. 

\begin{figure}[ht]
%\centerline{\includegraphics[width=0.5\textwidth]{Fig1hop0hop0_5.pdf}}
\centerline{\includegraphics[width=0.45\textwidth]{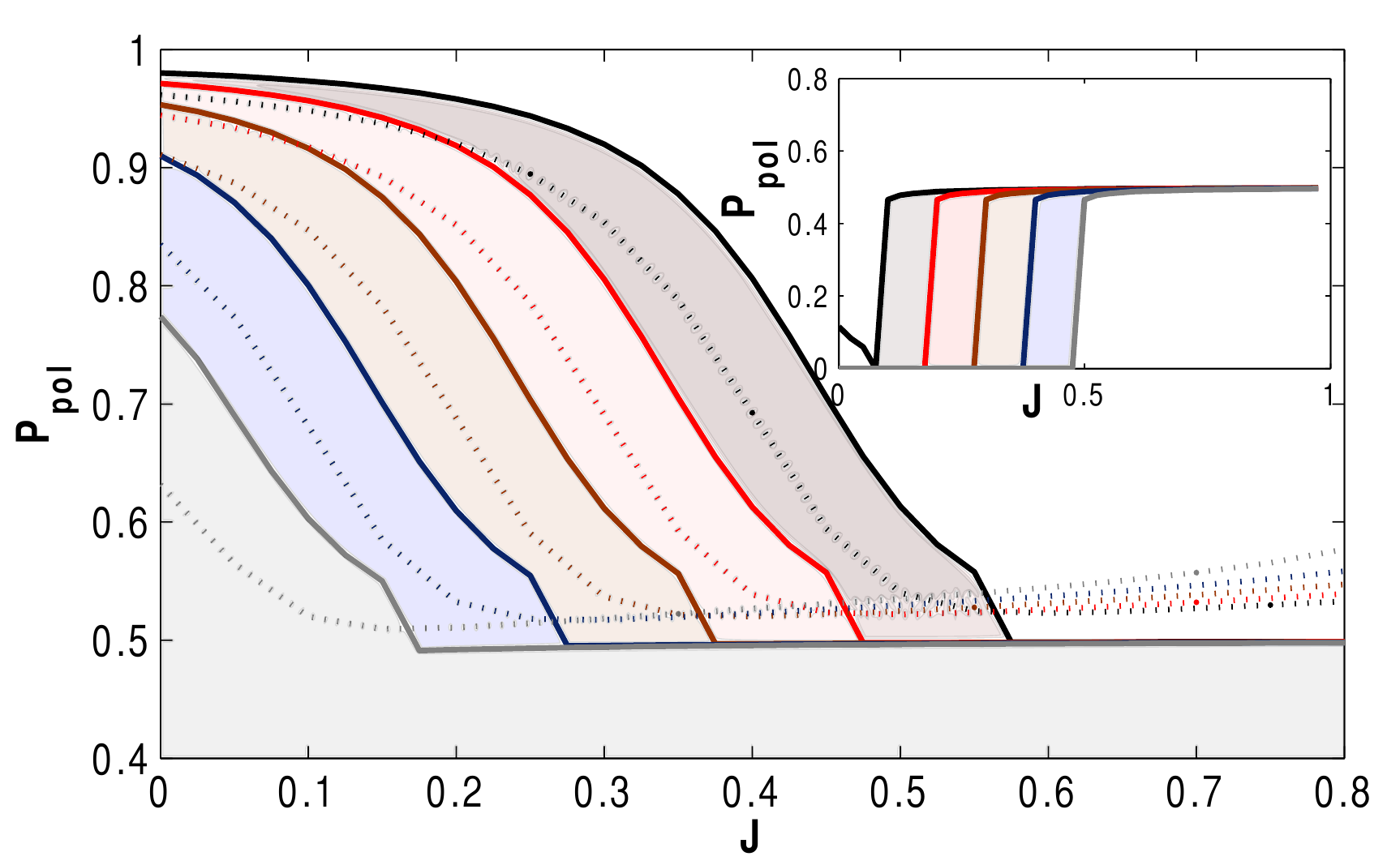}}
\caption{(Color online) Solid curves represent the total population in trapped in the form of polaritons $P_{\ppol}$, for different hopping rates $J$ and different atomic frequencies. The different curves correspond to values of $\omega_0$ ranging from $0.1$ (curve in the right extreme) to $0.5$ (curve in the left extreme) in an interval of $\Delta\omega_0=0.1$. These curves represent therefore values within the gap, while the inset represents $\omega_0$ within the band, ranging from $0.6$ (left extreme) to $1$ (right extreme). The dotted lines represent the total atomic population $P_T(t)$ at $t=500$ under the same conditions. In all curves we have considered $N=2$ atoms, $A=1$, $B=0.5$, and $g=0.1$. The initial condition is considered $|\Psi_0\rangle=|1\rangle|0\rangle|{\textmd vac}\rangle$, where $|{\textmd vac}\rangle$ is the vacuum state for the environment. See units in Sect. (\ref{units}).\label{polar2}}
\end{figure}

%
%\begin{figure}[ht]
%%\centerline{\includegraphics[width=0.45\textwidth]{PolaritonJvariosw.pdf}}
%\centerline{\includegraphics[width=0.45\textwidth]{JDecay.pdf}}
%%\centerline{\includegraphics[width=0.45\textwidth]{DosExcitTJ2.pdf}}
%%\centerline{\includegraphics[width=0.5\textwidth]{hop0hop0_5twoexcit.pdf}}
%\caption{Time evolution of the total population for $N=2$. The different panels correspond, from top to bottom, to atomic frequencies $\omega_0=0.1,0.25,0.5$. Each plot shows curves for different values of $J$, ranging from $J=0.5$ (low curve) to $J=0.95$ (top curve). Hence, higher values of $J$ lead to a slower decaying. \label{polar3}}
%\end{figure}

Let us now analyse further the formation of cavities inside the photonic crystal. To this order, we consider in figure (\ref{trapped}) a histogram of the atomic and photonic populations in the long time limit for two different situations: when the atomic frequency is $\omega_0=0.1$, i.e. deep inside the gap (left plots), and when it is just in the band-gap edge, $\omega_0=B=0.5$ (right plots). The two cases present very different results. While for for $\omega_0=0.1$ the population remains localized nearby the initially excited atom, for the resonant case $\omega_0=B=0.5$, the population spreads along the whole atomic sample, and along more modes within the chain. This result confirms numerically the formal discussion in \cite{john1990}, showing that indeed the localization length $\xi$ of a photon grows larger and eventually diverges near the band-gap edge $\xi\sim 1/(\sqrt{\omega_c|\omega_c-\omega_0|})$. Here, it can be seen that deep inside the gap, the excitation remains localized nearby the original location (see also \cite{devega2008c}), which effectively corresponds to the formation of a cavity where the cavity mode is just the transformed oscillator coupled to the atom. A similar result is observed for both cases when one and two excitations are initially present in the atomic lattice. 
\begin{figure}[ht]
%\centerline{\includegraphics[width=0.45\textwidth]{PolaritonJvariosw.pdf}}
\centerline{\includegraphics[width=0.5\textwidth]{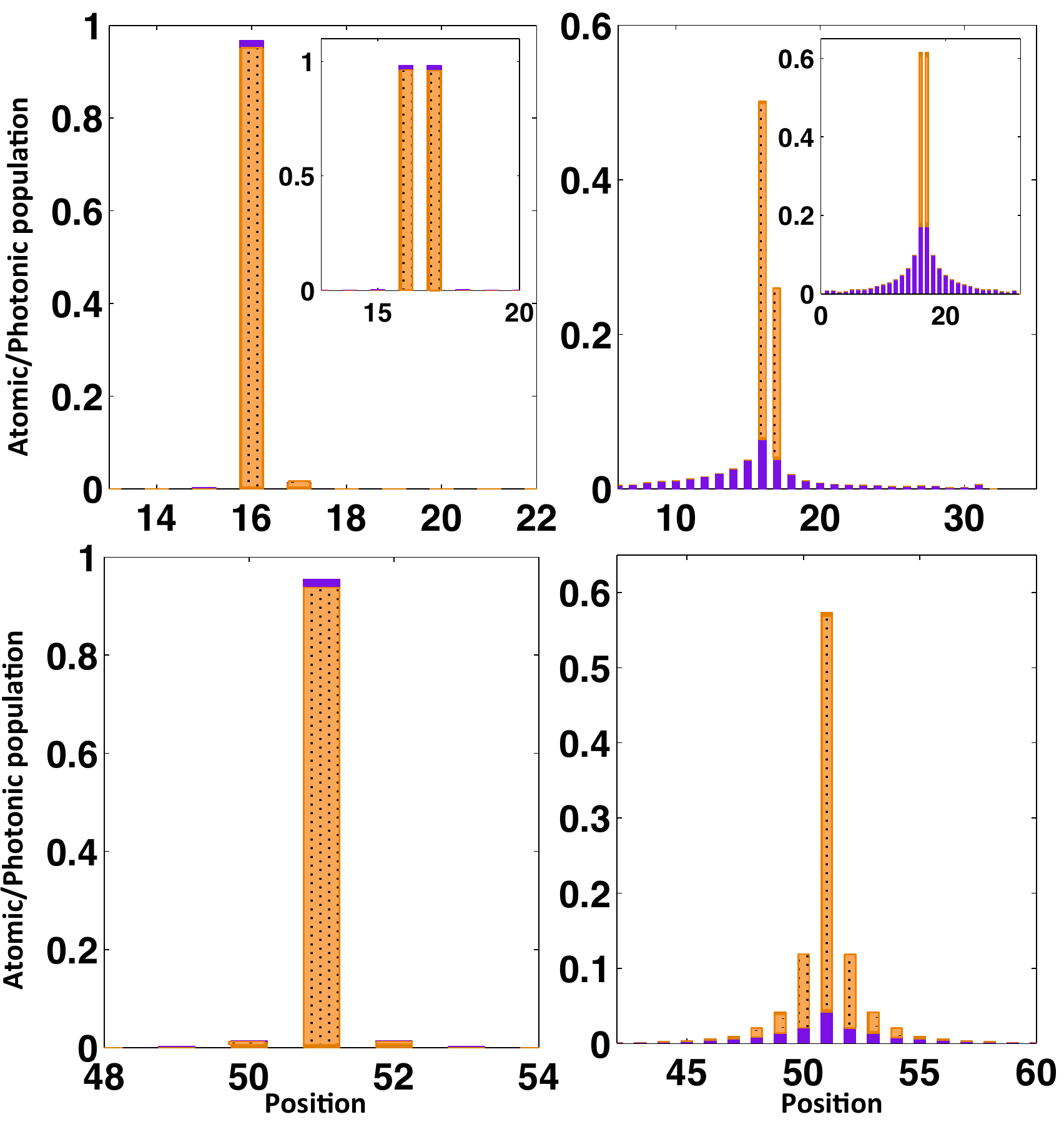}}
%\centerline{\includegraphics[width=0.45\textwidth]{DosExcitTJ2.pdf}}
%\centerline{\includegraphics[width=0.5\textwidth]{hop0hop0_5twoexcit.pdf}}
\caption{(Color online) Histograms representing the time-averaged atomic (dotted orange) and photonic (plain violet) populations within the long time limit. The left and right plots correspond to $\omega_0=0.1$ and $0.5$ respectively. The upper panels correspond to the case where $N=2$ atoms, considering one atom initially excited (and two for the insets). The lower panels represent the case of $N=15$ atoms, and a single initial excitation in the system. Other parameters considered are $M=50$, $B=0.5$ and $A=1$ as in previous cases. \label{trapped}}
\end{figure}

\section{Conclusions and perspectives}

The paper proposes a method to map a many body OQS conformed by a regular particle array into that of two coupled lattices, corresponding to the OQS and its environment respectively. The atoms within the OQS are then directly coupled to \textit{the most relevant modes of the environment}, which in principle may allow to solve the problem with a variety of techniques alternative to the master equation.
 %including the Schr{\"o}dinger equation, mean field, quantum Montecarlo, or DMRG. 
The mapping is particularly simple for atoms in contact with the electromagnetic field within a photonic crystal, that lead to interactions within the environment modes that extend only to next-neighbours. 
Generally speaking, in the case of bosonic environments, the mapping may lead to a system similar to the Jahn-Teller \cite{greentree2006,porras2012,nevado2012} and the Jaynes-Cummings for coupled cavities. 
%Finally, we have shown that: (a) truncating the environment chain beyond the nodes that are in direct constant with the many body system; and (b) considering the truncated part as a new reservoir, it is possible to describe the dynamics of the new system (many body system + coupled nodes) as a open quantum system, and to use the common Markov or weak coupling approximations. 
%Also, the problem of a strongly correlated spin chain coupled to a bosonic chain is well known in the literature as the Jahn-Teller model, and has been analysed in \cite{nevado2012,porras2012} in the framework of the mean field and spin-wave theories. 
%The most popular ones, master equations (ME) or stochastic Schr{\''o}dinger equations (SSE) \cite{wallsbook,breuerbook}, require often the weak coupling or the Markov approximations. Hence, a comparison between the results obtained from the analysis of the mapped system, and the original one would be extremely relevant to determine, for instance, the validity of such approximations.
Hence, re-expressing an OQS Hamiltonian as (\ref{fintrans}) allows to use both numerical and analytical tools alternative to the ones traditionally considered in the analysis of OQS, that can be helpful to analyse the system beyond the usual weak coupling and Markovian regimes. 

These ideas are illustrated in this paper by analysing the dynamics of atoms coupled to photonic crystals with a one dimensional gap. For this particular system, the result given by the master equation approach is compared to the result given by solving the Schr{\"o}dinger equation for the mapped system. Also, because the mapping allows to truncate to the most relevant modes of the environment, an exact diagonalization of the total Hamiltonian is carried out, which unveils the existence of highly correlated atom-photon states in the system when the atomic frequencies are within the gap, and a strong dependence of such highly correlated states on the presence of inter-atomic dipole-dipole interactions.

%We note that, since the analysis in this paper is based on an exact diagonalization and on a Schr{\"o}dinger equation for the total system, the number of excitations that can be considered is rather limited, although still gives a good illustration of the applications of the proposed transformation to deal with many body OQS. 
%Thus, it would be interesting to extend the study to many excitations, by considering more powerful numerical techniques, like t-DMRG or Montecarlo, or even approximated schemes like the mean field approach. Another interesting extension of the present work would be the study the dynamics of atoms, i.e. their absorption and emission properties, within two and three dimensional photonic band gap structures. 

Also, it is noted that the transformed system is very similar to the starting setup from which collision-model-based non-Markovian master equations are derived \cite{giovannetti2012,ciccarello2013}. These models give rise to master equations that preserve highly desirable mathematical properties such as complete positivity, but have the drawback that they are not easily derivable from microscopic models. For instance, the quantities involved in the collisional model approach cannot in general be expressed in terms of the environment spectral density. Also, collisional models cannot be easily extended to deal with a many body OQS. Hence, the chain-mapping approach here proposed could be used as a starting point to derive a new collision-model-based master equation that on the one hand preserves complete positivity, and on the other hand can be extended to the many particle case and be directly related to a microscopic derivation that departs from first principles, i.e. from the total Hamiltonian of the system and its environment. 

Being able to analyze the dynamics of many body OQS may bring new insight to a wide variety of problems, ranging from quantum optics (e.g. analysis of the dynamics of impurities in structured environments such as photonic crystals \cite{john1995,john1997}) and solid state physics (e.g. analysis of the dynamics of superconducting qubits \cite{you2005,you2012} and of strongly correlated systems with dissipation \cite{calzadilla2006,prosen2008,daley2009,schwanger2013,cai2013}), to quantum biophysics (e.g. study of the energy transport within photosyntetic complexes \cite{chin2013,rey2012}). In addition, understanding the dissipative dynamics beyond the Markov approximation is of primary importance to further develop the concepts of dissipative quantum computation and state preparation developed in \cite{verstraete2009,diehl2008,diehl2010}, and experimentally realized in \cite{subasi2012}. 
Finally, the appealing form of Hamiltonian (\ref{fintrans}) suggests that the dynamics of OQS can be simulated with optical lattices in a spirit similar to the proposals \cite{recati2005,devega2008}, provided that the interaction strengths $f_{nm}$ corresponding to a particular environment are implemented. An alternative implementation of this system, consisting in a regular lattice of atoms connected to an environment, are Coulomb crystals of trapped ions \cite{porras2008,mavadia2013}.

%{\bf Finally, the ideas here discussed pave the way for finding an orthonormal basis such that $\sum_k \phi^*_n(k)\phi_m(k)e^{-i\omega(k)t}$ decays with the distance $|n-m|$, what would allow us to deal with new physical situations, and to reproduce more realistically cooperative effects within three dimensional environments.} 
\textit{Acknowledgements}
The author gratefully acknowledges D. Alonso, M.C. Ba{\~n}uls, C. Busser, A. Gonzalez-Tudela, A. Perez, C. Zi and U. Schollw{\"o}ck for interesting discussions, and D. Alonso, J.I. Cirac, A. Ekert
%S.F. Huelga, M.B. Plenio 
and U. Schollw{\"o}ck for encouragement and support. This project was financially supported by the Nanosystems Initiative Munich (NIM) (project No. 862050-2) and partially by the Spanish MICINN (Grant No. FIS2010-19998).
\section*{Appendix A: System evolution equations and their dependency on the environment correlation function}

Let us now follow the derivation in \cite{alonso2007} to show that for initially thermal states, the only relevant quantity to describe the coupling with the environment is the so-called correlation function. To see this, we consider the simple case of an atom coupled to an environment with a Hamiltonian of the form (\ref{hEM1}). The idea is to consider the Heisenberg equation for a system operator, $A$, and re-express it in such a way that the environmental operators $a_{\bf k} (0)$ are placed on the right hand side of the terms, while the $a^\dagger_{\bf k} (0)$ appear in the left hand side. Thus, when computing $\langle A(t)\rangle=\Ttr[A(t)\rho_0]$, with $\rho_0=\rho_S\otimes\rho_T$, and $\rho_T$ a thermal state for the environment, these terms vanish.
The Heisenberg evolution equation of $A(t)={\mathcal U}^{-1}(t,0)A{\mathcal U}(t,0)$, with ${\mathcal U}(t,0)$ the evolution operator with the total Hamiltonian H, can be written as
\begin{eqnarray}
&&\frac{dA(t_1 )}{dt_1}=i{\mathcal U}^{-1}(t_1 ,0 )[H_{tot},A]{\mathcal U}(t_1 ,0 )\cr
&=&-i [H_S (t_1 ),A(t_1 )]+i\sum_{\bf k} g({\bf k}) \large( a_{\bf k}^{\dagger} (t_1,0 )[L(t_1 ),A(t_1 )]\cr
&+&[L^{\dagger}(t_1 ),A(t_1 )]a_{\bf k} (t_1 ,0 ) \large),
\label{eq39}
\end{eqnarray}
We can replace in (\ref{eq39}) the formal solution of the evolution equation of the environmental operators, $da_{\bf k} (t_1 ,0)/dt_1  =i[H_{tot}(t_1 ),a_{\bf k} (t_1,0 )]=-i\omega_{\bf k} a_{\bf k} (t_1,0 )-ig({\bf k}) L(t_1 )$,
\begin{eqnarray}
a_{\bf k} (t_1 ,0)&=&e^{-i\omega_{\bf k} t_1 }a_{\bf k}(0,0)-i g({\bf k}) \int^{t_1}_0 d\tau e^{-i\omega_{\bf k} (t_1-\tau)}\cr
&\times&L(\tau).
\label{eq40}
\end{eqnarray}
The single evolution equation (\ref{eq39}) becomes as follows,
\begin{eqnarray}
&&\frac{dA(t_1 )}{dt_1}=i [H_S (t_1 ),A(t_1 )]-\nu^{\dagger}(t_1 )[L(t_1 ),A(t_1 )]\nonumber \\
&+&\int_0^{t_1 }d\tau \alpha^*(t_1 -\tau) L^{\dagger}(\tau)[A(t_1 ),L(t_1 )]+[L^{\dagger}(t_1 ),A(t_1 )]\cr
&\times&\nu(t_1 )+\int_0^{t_1 }d\tau \alpha(t_1 -\tau)[L^{\dagger}(t_1 ),A(t_1 )]L(\tau),
\label{eq41}
\end{eqnarray}
where we have defined the environment correlation function as 
\begin{equation}
\alpha(t-\tau)=\sum_{\bf k} |g({\bf k})|^2 e^{-i \omega_{\bf k} (t-\tau)}.
\label{chapuno325}
\end{equation} 
In the last expression, we have also defined the bath operators
 \begin{eqnarray}
\nu^{\dagger}(t_1 )=-i\sum_{\bf k} g({\bf k})  a_{\bf k}^{\dagger} (0,0)e^{i\omega_{\bf k} t_1 }\nonumber  \\
\nu(t_1 )=i\sum_{\bf k} g({\bf k})  a_{\bf k} (0,0)e^{-i\omega_{\bf k} t_1}
\label{eq41b}
\end{eqnarray}
Note that when calculating the quantum mean value with an initial thermal state, the terms proportional to $\nu$ and $\nu^\dagger$ vanish, so that the exact evolution equation of $\langle A(t)\rangle$ only depends on the correlation function (\ref{chapuno325}). A similar calculation for two time correlation functions of system observables $A$ and $B$ leads to the form 
\begin{eqnarray}
&&\frac{dA(t_1 )B(t_2 )}{dt_1}=i [H_S (t_1 ),A(t_1 )]B(t_2 )
\cr&-&\nu^{\dagger}(t_1 )[L(t_1 ),A(t_1 )]B(t_2 )+[L^{\dagger}(t_1 ),A(t_1 )]B(t_2 )\nu(t_1 )\nonumber\\
&-&\int_0^{t_1 }d\tau \alpha^* (t_1 -\tau)L^{\dagger}(\tau)[L(t_1 ),A(t_1 )]B(t_2 )\nonumber\\
&+&\int_{t_2}^{t_1 }d\tau \alpha(t_1 -\tau)[L^{\dagger}(t_1 ),A(t_1 )]L(\tau)B(t_2 )\cr
&+&\int_0^{t_2} d\tau  \alpha(t_1 -\tau)[L^{\dagger}(t_1 ),A(t_1 )]B(t_2 )L(\tau).
\label{eq45}
\end{eqnarray}
The evolution of the quantum mean value $\langle A(t_1 )B(t_2 )\rangle$ is again obtained by computing the trace with the total initial state on both sides of the former expression, finding out that the resulting exact equation only depends on the correlation function. A generalization to an N-time correlation function can be found in \cite{alonso2007}.

\section*{Appendix B: Independent environment limit}
To illustrate the idea presented in Section (\ref{indep}), let us assume the Caldeira and Legget model for the spectral density, which gives a good approximation of the spectral densities of different types of environment in the short frequency limit  \cite{caldeira1983,wallsbook,leggett1987},
\begin{eqnarray}
J(\omega)=\alpha\omega_c^{1-s}\omega^s\theta(\omega_c-\omega).
\label{chapuno41}
\end{eqnarray}
Here, $0<s<1$ in the \textit{sub-ohmic} case, and $s>1$ in the \textit{super-ohmic} 
 where $\omega_c$ is a frequency cut. To reproduce such spectral density, we assume a dispersion relation of the form $\omega(k)=Ak^p$, where $A$ and $p$ are constants to be chosen as convenient. In terms of this, the corresponding density of states is $\rho_\DDOS(\omega)=\large|\frac{\partial k}{\partial \omega}\large|=\frac{A^{\frac{1-p}{p}}}{p}\omega^\frac{1-p}{p}$. Considering $g(k)=g$, i.e. a homogeneous coupling Hamiltonian as in (\ref{h1EM}), we find that
\bea 
J(\omega)=g^2\rho_\DDOS(\omega)=g^2\frac{A^{\frac{1-p}{p}}}{p}\omega^\frac{1-p}{p}.
\eea
%\begin{figure}[ht]
%%\centerline{\includegraphics[width=0.5\textwidth]{chain2.pdf}}
%%\centerline{\includegraphics[width=0.5\textwidth]{tresH.pdf}}
%%\centerline{\includegraphics[width=0.5\textwidth]{ChainOnly.pdf}}
%%
%\centerline{\includegraphics[width=0.45\textwidth]{Refnm.pdf}}\centerline{\includegraphics[width=0.45\textwidth]{ThreeDCaldeira.pdf}}
%\caption{Decaying of the coefficient $f_{np}$ with $p$ for $n=1$ (curves for $n\neq 1$ have a maximum in $m=n$ and the same form). Circles and squares correspond to the ohmic and sub-ohmic spectral densities ($q=1/2$ and $q=2/3$ respectively). Diamonds and triangles correspond, respectively, to the real and imaginary part of $f_{1p}$ for super-ohmic spectral densities ($q=1/3$). \label{caldero}}
%\end{figure}
Hence, to reproduce an ohmic spectral density ($s=1$), we need to chose $p=1/2$, so that $(1-p)/p=1$, and the constant $A=1/2$. Similarly, a sub-ohmic spectral density of the form ($s=1/2$) will require choosing $p=2/3$ and $A=4\omega_c/9$, and a super-ohmic like $J(\omega)=\frac{\omega^{2}}{\omega_c}$ will require $p=1/3$ and $A=\sqrt{\frac{1}{3\omega_c}}$. In all cases, the coupling should be chosen as $g=\sqrt{\alpha}$.
In general, for a spectral density with $s$, we need $p=1/(s+1)$ and $A=(\frac{\omega_c^{1-s}}{s+1})^{1/s}$. 
%Fig (\ref{caldero}) represents the decaying of the coefficients (\ref{coefmap}) for the needed value ${\bf r}_l=0$, and considering ${\bf r}_n=(1,0,0)d_0$.

%Also, the ability to understand energy transport in open quantum systems subject to noise and dissipation beyond the Markov limit may lead to important insight in the analysis of photosynthetic systems, and the design of nano-estructured solar cells where such energy transport can actually be controlled and directed to specific parts of the cell. 

%\bibliography{/Users/inesdevega/Documents/Trabajo/Referencias/Bibtexeles}
\bibliography{/Users/ines.vega/Dropbox/Bibtexelesdrop}
%\bibliography{/Users/inesdevega/Dropbox/Bibtexelesdrop}
\bibliographystyle{prsty}
%\bibliography{lattice}
%\bibliographystyle{unsrt}
%\begin{thebibliography}{99}
%%%%%%%%%%%%%%%%%%%%%%%%%%%%%%%%%%%%%%%%%%%%%%%%%%%%%%%%%%%%
%\bibitem{devega08} I. de Vega \textit{et al.}, Phys. Rev. Lett. 101, 260404 (2008).
%\bibitem{recati04} A. Recati \textit{et al.}, Phys. Rev. Lett. 94, 040404 (2005).
%\end{thebibliography}

\end{document}